\documentclass{aastex}

\usepackage{graphicx}
\usepackage{emulateapj5}
\usepackage{times}



\def\chandra    {{\em Chandra}\/}

\def\xmm        {XMM-{\em Newton}\/}
\def\rosat      {{\em ROSAT}\/}
\def\hst        {{\em HST}\/}
\def\spi        {{\em Spitzer}\/}

\def\ga         {{ESO~137-006}\/}

\usepackage{mathptmx}

\begin{document}

\title{Every BCG with a strong radio AGN has an X-ray cool core: \\is the cool core --- noncool core dichotomy too simple?}

\author{
M.\ Sun
}
\smallskip

\affil{Department of Astronomy, University of Virginia, P.O. Box 400325, Charlottesville, VA 22904-4325; msun@virginia.edu}

\shorttitle{Radio AGN and cool cores}
\shortauthors{Sun}

\begin{abstract}

Radio AGN feedback in X-ray cool cores has been proposed as a crucial ingredient
in the evolution of baryonic structures. However, it has long been
known that strong radio AGN also exist in ``noncool core'' clusters,
which brings up the question whether an X-ray cool core is always required
for radio feedback. In this work, we present a systematic analysis of
BCGs and strong radio AGN in 152 groups and clusters from the \chandra\
archive. All 69 BCGs with radio AGN more luminous than
2$\times10^{23}$ W Hz$^{-1}$ at 1.4 GHz are found to have X-ray cool cores.
The BCG cool cores can be divided into two classes, the
large-cool-core (LCC) class and the corona class. Small coronae, easily
overlooked at $z>$0.1, can trigger strong heating episodes in groups and
clusters, long before large cool cores are formed. Strong radio outbursts
triggered by coronae
may destroy embryonic large cool cores and thus provide another mechanism
to prevent formation of large cool cores. However, it is unclear whether
coronae are decoupled from the radio feedback cycles as they have to be largely
immune to strong radio outbursts. Our sample study also shows the absence of groups
with a luminous cool core while hosting a strong radio AGN, which is not observed in clusters.
This points to a greater impact of radio heating on low-mass systems than clusters.
Few $L_{\rm 1.4GHz} > 10^{24}$ W Hz$^{-1}$ radio AGN ($\sim$ 16\%) host a
$L_{\rm 0.5 - 10 keV} > 10^{42}$ ergs s$^{-1}$ X-ray AGN, while above these
thresholds, all X-ray AGN in BCGs are also radio AGN.
As examples of the corona class, we also present detailed analyses of a BCG corona associated with
a strong radio AGN (\ga\ in A3627) and one of the faintest coronae known
(NGC~4709 in the Centaurus cluster).
Our results suggest that the traditional
cool core / noncool core dichotomy is too simple. A better alternative is the
cool core distribution function, with the enclosed X-ray luminosity or gas mass.

\end{abstract}


\section{Introduction}

The importance of AGN outflows for cosmic structure formation and evolution
has been widely appreciated recently. AGN outflows may simultaneously explain
the antihierarchical quenching of star formation in massive galaxies, the
exponential cut-off at the bright end of the galaxy luminosity function (LF),
the $M_{\rm SMBH} - M_{\rm bulge}$ relation and the quenching of cooling-flows
in cluster cores (e.g., Scannapieco et al. 2005; Begelman \& Nath 2005;
Croton et al. 2006; Best et al. 2006). Outflows from radio AGN are especially important locally
because nearly all strong radio AGN are hosted by early-type galaxies that
dominate the high end of the LF and the galaxy population in clusters.
Radio AGN are different from emission-line AGN selected in optical surveys.
Kauffmann et al. (2004) showed that optical emission-line AGN favor low-density
environments and are mainly produced by BHs with masses below 10$^{8}$ M$_{\odot}$.
On the other hand, radio-loud AGN favor denser environments
than do normal galaxies, and the integrated radio luminosity density comes from
the most massive BHs (Best et al. 2005). Best et al. (2005) also showed that
the fraction of galaxies that are radio AGN increases with the stellar or BH mass
(as $M_{*}^{2.5}$ or $M_{BH}^{1.6}$), while Shabala et al. (2008) found a similar
relation as $M_{*}^{2.1\pm0.3}$ or $M_{BH}^{1.8\pm0.5}$.
The analysis by Best et al. indicates no correlation between radio and optical
emission-line luminosities for radio AGN, and the probability that a galaxy
of given mass is radio loud is independent of whether it is an optical AGN.
These results drove Best et al. (2005) to suggest that low-luminosity
radio AGN (FR I or fainter) form a distinctly different group from
optical emission-line AGN. They suggested that the radio AGN activity is associated
with cooling of gas from the hot halos surrounding elliptical galaxies and
clusters (see also Hardcastle et al. 2007).
A similar idea has been proposed by Croton et al. (2006). Besides
the classical ``quasar mode'' in AGN feedback, they introduced a ``radio
mode'' which is the result of the X-ray gas accretion on to the central
BHs. The inclusion of this ``radio mode'' in their simulations
allows suppression of both excessive cooling and growing of very massive
galaxies. It also explains why most massive galaxies are red
bulge-dominated systems containing old stars.

Radio activity of galaxies is enhanced in clusters and the distribution of
cluster radio AGN is highly concentrated (e.g., Lin \& Mohr 2007;
Best et al. 2007). Among all cluster galaxies,
the brightest cluster galaxy (BCG) is generally the most massive galaxy and
has the biggest impact on the surrounding large cool core and the general ICM properties.
With the SDSS C4 cluster sample, Best et al. (2007) found that BCGs are
more likely to host a radio AGN than other galaxies of the same stellar
or BH mass (see also Lin \& Mohr 2007). This difference implies that either
BCGs stay in each radio active cycle for a longer time, or they are more
frequently re-triggered than non-BCGs. The mutual interaction between radio AGN
(especially those of BCGs) and the surrounding ICM has
a significant impact on both of them. The great
power of jets from radio AGN can not only quench cooling in cluster cool
cores, but also can drive the ICM properties away from those defined by simple
self-similar relations involving only gravity (summarized in Voit 2005). The
ICM around the radio AGN provides a historical chronicle of the SMBH activity.
X-ray cavities and shocks serve as calorimeters for the total energy outputs
of AGN that allow us to understand a great deal of AGN feedback and SMBH
growth (summarized by McNamara \& Nulsen 2007). 
On the other hand, Radio AGN (FR-I or fainter) may need an enhanced X-ray
atmosphere to fuel them.
Churazov et al. (2005) used Galactic X-ray binaries as analogue for the
different evolution states of the SMBH, from radiative efficient mode with weak
outflows (quasars) to radiative inefficient mode with strong outflows (local giant
ellipticals). The SMBHs of local giant ellipticals have very low accretion rates and
radiative efficiencies, but are very efficient to heat the surrounding gas.
Allen et al. (2006) presented a tight correlation between the Bondi accretion rate
and the mechanical power of radio AGN for BCGs in 9 groups and clusters.
The relation implies that 1\% - 4\% of mass energy of the gas accreted through
Bondi accretion is transferred to the feedback energy to heat the surrounding
ICM. Hardcastle et al. (2007) further suggested that accretion of hot gas is
sufficient to power all low-excitation radio AGN (FR-I and some FR-II), while the
high-excitation FR-II AGN are powered by accretion of cold gas.

As $\dot{M}_{Bondi} \propto M_{BH}^{2} K^{-1.5}$ ($K$ is entropy of the surrounding gas
defined as $kT / n_{e}^{2/3}$), strong radio AGN require low-entropy circum-nuclear
gas and massive black holes. Although many groups and clusters have large and dense cool
cores (e.g., $\sim$ 50\% in the HIFLUGCS sample, Chen et al. 2007), most cluster galaxies
(including many BCGs) are not located in such large cool cores. Their surrounding ICM has
too high an entropy ($>$ 50 times the value at the center of cluster cool cores)
to trigger any significant nuclear activity.
Thus, the emerging significant questions are: {\em What fuels the strong radio AGN in
groups and clusters, especially those not in large cool cores? Is an X-ray cool
core always required for strong radio AGN in groups and clusters? Is the sufficient
hot gas supply the reason that BCGs stay in the radio active phase longer?}
The work mentioned in the last two paragraphs only discussed optical and radio data
of radio AGN. \chandra\ and \xmm\ allow detailed studies of the X-ray atmosphere
around radio AGN in groups and clusters. Small X-ray thermal halos around the nuclei
of strong radio AGN are often found (e.g., Hardcastle et al. 2001 on 3C~66B;
Hardcastle et al. 2002 on 3C~31; Worrall et al. 2003 on NGC~315; Hardcastle et al. 2005
on 3C296; Sun et al. 2005b on NGC~1265; Evans et al. 2006 on a sample of FR-I and FR-II
galaxies). Hardcastle et al. (2007) also summarized some recent \chandra\ and \xmm\
results for the hot gas atmosphere of nearby radio galaxies but the sample with detailed
studies is small and those radio AGN are mainly in groups. 

With different original motivations,
Sun et al. (2007, S07 hereafter) presented a systematic analysis of X-ray thermal
coronae of early-type galaxies in 25 hot ($kT>$3 keV), nearby ($z<$0.05) clusters,
based on \chandra\ archival data. Small and cool galactic coronae ($\leq$ 4 kpc
in radius and $kT$=0.5-1.1 keV generally) have been found to be common, $>60$\% in
$L_{\rm Ks}>$2 $L*$ galaxies. Although much smaller and fainter than the classical
cluster cool cores like those in Perseus, Virgo and A478, they are in fact mini-versions
of large cool cores (with cooling time of 10 Myr - 1 Gyr from the center to the boundary),
composed of the ISM gas (from the stellar mass loss) pressure confined by the surrounding ICM.
These small coronae managed to survive strong ICM stripping, evaporation, rapid cooling, and
powerful AGN outflows (see S07 for detailed discussions). One interesting result noticed by S07
is the connection between strong radio AGN and small coronae (Section 4.4 of S07).
However, the S07 sample size is small (9 galaxies with $L_{\rm 1.4 GHz} > 10^{24}$ W Hz$^{-1}$)
and only small coronae were discussed.
Sun et al. (2009, S09 hereafter) added six more examples of BCG coronae with strong radio AGN but
the combined sample with S07 is still small.
The questions raised early need to be addressed with a bigger sample.
In this work, we present results on such a sample from the \chandra\ archive
(186 galaxies from 152 groups and clusters, including all BCGs, 74 galaxies with
$L_{\rm 1.4 GHz} > 10^{24}$ W Hz$^{-1}$). The plan of this paper is as follows:
The galaxy sample is defined in Section 2. The data analysis is presented
in Section 3, as well as the definition of cool cores in this work.
In Section 4, we discuss the X-ray gas atmosphere of BCGs and non-BCGs
with luminous radio AGN from the sample study. After learning the general properties
of the sample, we present a detailed analysis of a luminous corona associated
with a strong radio AGN in Section 5 (\ga). Many \chandra\ data are shallow
so only upper limits of coronal emission exist in some cases. In Section 6, we discuss the
the faintest corona known that sheds light on the properties of faint embedded coronae. 
Discussions are in Section 7 and Section 8 contains the conclusions.
We assumed H$_{0}$ = 71 km s$^{-1}$ Mpc$^{-1}$, $\Omega$$_{\rm M}$=0.27,
and $\Omega_{\rm \Lambda}$=0.73.

\section{The sample}

We want to examine the X-ray gas component associated with BCGs and other strong radio
AGN, either large cool cores or small coronae.
The typically small luminosity of a corona ($\sim 10^{41}$ ergs s$^{-1}$ in the
0.5 - 2 keV band) limits our studies to local systems. S07 only studied
$kT > 3$ keV clusters at $z<0.05$. In this work, we include any \chandra\
observations of $z<0.065$ groups and clusters with an exposure of longer than
5 ks. We also include higher redshift systems with sufficient \chandra\ data,
but with a hard limit of $z\leq 0.11$. As we want to cover more volume in a
single group or cluster, we limit our sample to $z>0.01$ systems.
There are very few strong radio AGN in $z<0.01$ groups and clusters anyway.
The group and cluster sample is collected from the \chandra\ archive. Two X-ray
flux-limited samples, B55 (Peres et al. 1998) and HIFLUGCS (Reiprich \& B$\ddot{\rm o}$hringer
2002), are almost 100\% covered by \chandra. However, both samples
are not big and have few or no groups. As less than 1/3 of the REFLEX and NORAS clusters
(B$\ddot{\rm o}$hringer et al. 2000; B$\ddot{\rm o}$hringer et al. 2004) have \chandra\
data, we are forced to include as many systems from the archive as we can. The final
sample includes 152 groups and clusters (Table 1). Most conclusions of this work
should not be much affected by the heterogeneous nature of the cluster sample. We will
also discuss the B55 and the extended HIFLUGCS samples in Section 7.4.

Our main goal is to examine the X-ray gas atmosphere around BCGs so all BCGs are
included in the galaxy sample. The BCG is defined as the most luminous galaxy
in the 2MASS $K_{\rm s}$ band within 0.1 $r_{500}$ of the group or cluster.
In relaxed groups and clusters, the selection of BCGs is easy. In nine unrelaxed
groups and clusters, two BCGs with comparable $K_{\rm s}$-band luminosities are
selected, including Coma, A2147, A1367, A1142, A3128, RXC J1022.0+3830, SC1329-313,
A1275 and A2384. Therefore, the BCG sample includes 161 galaxies.
We also want to include all galaxies with a 1.4 GHz luminosity higher than
$10^{24}$ W Hz$^{-1}$ ($L_{\rm 1.4 GHz, cut}$).
A radio spectral index of -0.8 is assumed in the small radio $K$-correction.
This $L_{\rm 1.4 GHz, cut}$ is about 1/3 of the $L_{*}$ of the radio luminosity
function (Best et al. 2005) and is comparable to the luminosities of many nearby
3C or PKS radio galaxies. For comparison, the central radio galaxy of the Centaurus cluster
(NGC~4696, PKS~1245-41) has an $L_{\rm 1.4 GHz}$ of 0.60$\times 10^{24}$ W Hz$^{-1}$. 
The giant radio galaxy in our backyard, Centaurus A, has an $L_{\rm 1.4 GHz}$ of
0.46$\times 10^{24}$ W Hz$^{-1}$.
The chosen $L_{\rm 1.4 GHz, cut}$ corresponds to an average mechanical power of
$\sim 6\times10^{43}$ ergs s$^{-1}$ (but with large scatter, B\^{i}rzan et al. 2008),
which is capable of increasing the thermal energy of a 10$^{11}$ M$_{\odot}$ gas
core (typical for the cool cores of luminous groups and poor clusters) at $kT$=1 keV
by $\sim$ 40\% in 100 Myr. For a small corona with a total gas mass of $\sim 10^{8}$ M$_{\odot}$
(S07), this kind of radio outbursts will easily destroy the whole corona if only
$\sim$ 1\% of the total mechanical power is deposited within the corona for 10$^{8}$ yr.
The radio fluxes come from the NRAO VLA Sky Survey (NVSS) (Condon et al. 1998)
and the Sydney University Molonglo Sky Survey (SUMSS) (Bock, Large \& Sadler 1999).
We also examine the higher resolution images from the Faint Images of the Radio
Sky at Twenty Centimeters (FIRST) survey (Becker, White \& Helfand 1995) and
literature if available. The origins of radio sources can be
determined in all cases. Because of the proximity, spectroscopic redshifts
are available for all radio galaxies in our sample. 
We generally use the system redshift (Table 1)
to calculate distance. For the Centaurus cluster and the NGC~7619 group,
we used their surface-brightness-fluctuation distance from Tonry et al. (2001).

Radio AGN are over-represented in our BCG sample. There are 50\% of BCGs
(81 out of 161) with $L_{\rm 1.4 GHz} > 10^{23}$ W Hz$^{-1}$ and 32\%
of BCGs (52 out of 161) with $L_{\rm 1.4 GHz} > 10^{24}$ W Hz$^{-1}$. Lin \& Mohr
(2007) derived the radio active fractions of BCGs for 573 groups and clusters selected
from the NORAS and REFLEX cluster catalogs. Their fractions at these two thresholds are
33\% and 20\% respectively. von der Linden et al. (2007) and Best et al. (2007)
examined a sample of 1106 groups and clusters optically selected from SDSS.
The radio active fraction of BCGs is $\sim$ 30\% at $L_{\rm 1.4 GHz} > 10^{23}$ W Hz$^{-1}$,
for the typical stellar mass of BCGs in our sample.
This fact again demonstrates the heterogeneous nature of our sample, which should be kept in
mind when our results are interpreted.
In fact, radio AGN are also over-represented in the B55 sample and the extended HIFLUGCS sample
(Section 7.4). The radio AGN fractions of BCGs in these two samples are similar to
ours, $\sim$ 50\% for $L_{\rm 1.4 GHz} > 10^{23}$ W Hz$^{-1}$ AGN and $\sim$ 30\% for
$L_{\rm 1.4 GHz} > 10^{24}$ W Hz$^{-1}$ AGN.
As X-ray flux-limited samples, the over-abundance of radio AGN of BCGs in both samples
is mainly caused by the prevalence of large cool core clusters in both samples, as strong radio
AGN are more likely to be associated with luminous cool core clusters (see Section 4).

\section{The \chandra\ data analysis and the cool core definition}

All observations were performed with the \chandra\ Advanced CCD Imaging Spectrometer (ACIS).
Standard data analysis was
performed which includes the corrections for the slow gain change
\footnote{http://cxc.harvard.edu/contrib/alexey/tgain/tgain.html}
and charge transfer inefficiency (for both the FI and BI chips). 
We investigated the light curve of source-free regions (or regions
with a small fraction of the source emission) to identify and exclude
time intervals with strong particle background flares.
We corrected for the ACIS low-energy quantum efficiency (QE) degradation
due to the contamination on ACIS's optical blocking filter 
\footnote{http://cxc.harvard.edu/cal/Acis/Cal\_prods/qeDeg/index.html},
which increases with time and is positionally dependent. The dead area
effect on the FI chips, caused by cosmic rays, has also been corrected.
We used CIAO3.4 for the data analysis. The calibration files used
correspond to \chandra\ calibration database (CALDB) 3.5.2 from the
\chandra\ X-ray Center. In the spectral analysis, a lower energy cut
of 0.4 keV is used to minimize the effects of calibration uncertainties
at low energy. The solar photospheric abundance table by Anders \& Grevesse
(1989) is used in the spectral fits. In the spectral analysis, cash statistics
is used for faint sources. Uncertainties quoted in this paper
are 1 $\sigma$. Special data analysis related to \ga\ is discussed in Section 5.1.

In this work, X-ray cool cores include both large cool cores and
small coronae. How do we define a large cool core? A core is considered as
a large cool core if its central isochoric cooling time is
less than 2 Gyr. The X-ray luminosity of a large cool core is measured within
a radius where the gas isochoric cooling time is 4 Gyr ($r_{\rm 4 Gyr}$).
The cooling time profiles come from S09, Cavagnolo et al. (2009)
\footnote{http://www.pa.msu.edu/astro/MC2/accept/}
and our own work.
Note that $r_{\rm 4 Gyr}$ effectively means the whole region for small coronae (S07 and Section 5).
Reasons behind these thresholds of cooling time are as follows. First,
there are systems with central cooling time between 1 Gyr and 10 Gyr,
or weak-cool-core clusters (e.g., Mittal et al. 2009; Pratt et al. 2009; Cavagnolo et al. 2009).
Some weak cool cores also have a BCG corona, e.g., A3558 with an ICM central
cooling time of $\sim$ 4 Gyr (S07), A1060 with an ICM central cooling time of
$\sim$ 3.6 Gyr (Yamasaki et al. 2002) and A2589 with an ICM central cooling time of
$\sim$ 2.5 Gyr beyond the central source (this work). We want to select a cut of
the central cooling time that excludes these sources so 2 Gyr is chosen.
Systems with central cooling time of $<$ 1 Gyr are usually considered as
strong cool cores (e.g., Mittal et al. 2009; Pratt et al. 2009). Our threshold
of 2 Gyr includes some weak cool cores, but not many. There are
seven clusters in this sample with a central cooling time of 1 - 2 Gyr,
A3571, A2244, A2063, A2142, A4038, A1650 and A2384 (north). None of them
has a BCG corona (likely already merged with the large ICM cool core) and their
radio AGN are always faint ($L_{\rm 1.4 GHz} < 6\times10^{22}$ W Hz$^{-1}$).
Thus, all large cool cores with $L_{\rm 1.4 GHz} > 2\times10^{23}$ W Hz$^{-1}$ in this
work are strong cool cores with a central cooling time of $\leq$ 1 Gyr (see Section 4).
Second, we want these two thresholds not too close to reflect the still high luminosities
of the weak cool cores, but also not too far to have many
systems fall between two thresholds. Thus, 4 Gyr is chosen as the aperture for
the luminosity measurement. A somewhat different cooling time (3 - 5 Gyr)
does not affect any conclusions of this work. There are seven systems in this ``grey'' area
with a central cooling time of 2 - 4 Gyr, including A2589, A2657, A3558, A1060,
AS405, A3266 and A3562. Based on our definition, they are not large cool cores
so any cool cores associated with their BCGs can only be small coronae, which
indeed is the case for some of them (A3558, A1060 and A2589). AS405 has the most
luminous radio AGN in this group with a $L_{\rm 1.4 GHz}$ of 1.65$\times10^{23}$ W Hz$^{-1}$,
while the radio AGN with other BCGs are all fainter than 3$\times10^{22}$ W Hz$^{-1}$.
As our focus is on BCGs with strong radio AGN, the exact classification of these
systems in the ``grey'' area have little impact on our conclusions. In fact, we can
conclude that weak cool cores do not have strong radio AGN, unless a corona is
present.

Section 3 of S07 detailed the analysis on the identification of faint thermal sources.
We always perform a spectral analysis. Most sources studied in this work have
sufficient counts for a detailed spectral analysis. However, there are faint sources
(e.g., with less than 100 counts in the 0.5 - 2 keV band) so the significance
of the iron L-shell hump or the ``softness'' of the spectrum needs to be tested (see Section 3
of S07). We used the S07 method with Monte-Carlo simulations (Section 3.1 of S07)
to address the significance of the iron L-shell hump. Only sources with an iron
L-shell hump that is more significant at $>$ 99.5\% level are considered
thermal coronae. For faint sources that do not meet the above criteria, we
further identified ``soft X-ray sources'' with a power law fit as defined in
Section 3.2 of S07. The power law index method is similar to the 1.5 - 7 keV / 0.5 - 1.5 keV
hardness ratio method, as coronae should have strong excess emission above the continuum
in the 0.5 - 1.5 keV band.
As argued in S07, most ``soft X-ray sources'' identified by the power law index method
should be genuine coronae. In fact, after the S07 work, deeper \chandra\ data
for A3627 and A2052 had been available. The data cover three sources identified as
``soft X-ray sources'' by S07, ESO~137-008 in A3627, CGCG~049-092 and PGC~093473
in A2052 (Table 2 of S07). The deeper data clearly reveal a significant iron L-shell hump
in each spectrum and confirm the corona nature for all of them.
This work identified 73 small coronae. Only seven of them are flagged as ``soft X-ray sources''
that are not as robust as others (five in Table 2). Thus, this uncertainty has
little impact on the conclusions of this work.
For sources not identified as coronae or ``soft X-ray sources'', upper limits on the
thermal emission are given, with the method listed in S07.

We emphasize that the number of coronae determined in this work is only a lower limit,
as \chandra\ exposures are not optimized for studies of faint galactic emission.
As shown in Section 6, the faintest embedded corona known has
an X-ray luminosity much lower than any upper limits in this work.
About 2/3 of galaxies only with upper limits of coronal emission are detected in
X-rays. They are either very faint sources or very bright X-ray AGN that makes the
identification of thermal emission difficult. NGC~3862 (or 3C~264) in A1367 is
a good example (S07, also in Table 2). If not for an on-axis sub-array observation, the weak
thermal emission of its corona cannot be confirmed only from the spectral analysis.
Different from S07, we also did not use any stacking and always studied the spectrum
of a single galaxy. Thus, the identification of thermal coronae in this work was
conservative.

The main cool-core property we measured is the 0.5 - 2 keV luminosity. This energy band is
where the thermal emission of a corona is significant, even if a strong X-ray AGN is present.
For large cool cores, the luminosity is derived from the spectral fit to the total
spectrum within $r_{\rm 4 Gyr}$. Multi-$kT$ components are often required for bright cores.
For coronae, the immediate local background was always used. A power-law component was
always included in the spectral fits to account for the nuclear and LMXB emission.
The gas abundance cannot be constrained for faint coronae so it was fixed at 0.8 solar
as derived by S07
\footnote{This assumed abundance corresponds to 1.3 solar in the new solar abundance table
by Lodders (2003).}.
The derived typical abundance for bright coronae is also consistent with
this assumed value (e.g., see Section 5.3 for \ga). 
For faint coronae, the uncertainty of the X-ray luminosity mainly comes from the
statistical error. The uncertainty of the non-thermal component does not affect the
0.5 - 2 keV luminosity much, as the identified faint coronae have dominant thermal
emission in the soft band. Faint sources with a hard spectrum will not be identified as
coronae as stated previously in this section.

\section{Radio AGN and X-ray cool cores}

We first examine BCGs in the $L_{\rm 0.5-2 keV}$ ($r<r_{\rm 4 Gyr}$) - $L_{\rm 1.4 GHz}$
plane (Fig. 1). BCGs in groups ($kT <$ 2 keV), poor clusters (2 keV $< kT <$ 4 keV)
and rich clusters ($kT >$ 4 keV) are color-coded. The system temperatures
come from BAX\footnote{http://bax.ast.obs-mip.fr/}, S09,
Cavagnolo et al. (2009) and our own work. As shown in Fig. 1, almost all
cool cores (including upper limits) can be divided into two classes, marked by a
vertical ellipse for small coronae and a tilted ellipse for large cool cores.
The dividing line between the two classes is $L_{\rm 0.5-2 keV} \sim
4\times10^{41}$ ergs s$^{-1}$. The gap between two classes is especially
significant at $L_{\rm 1.4 GHz} > 2\times10^{23}$ W Hz$^{-1}$.
There are 52 radio sources with $L_{\rm 1.4 GHz} > 10^{24}$ W Hz$^{-1}$ and 81
radio sources with $L_{\rm 1.4 GHz} > 10^{23}$ W Hz$^{-1}$.
Above $L_{\rm 1.4 GHz} > 2\times10^{23}$ W Hz$^{-1}$, {\bf every} BCG has a
confirmed cool core, small or large. In fact, there are only two non-detections of
cool cores out of 81 BCGs above $L_{\rm 1.4GHz}$ of $10^{23}$ W Hz$^{-1}$, which is the
dividing line between the star-formation and AGN components in the local radio LF
(e.g., Sadler et al. 2002). Below that threshold, radio emission
from star formation begins to dominate the local radio LF. This threshold
was also used in the statistical studies by Lin \& Mohr (2007) and von der Linden et al. (2007).
Upper limits of both non-detections (AS405 and A2572) are high (Fig. 1) as both
observations are short (7.9 ks) with ACIS-I, especially for AS405 at $z$=0.0613.
A faint soft X-ray point source is actually detected in the position of A2572's BCG ($z$=0.0403)
but was rejected as a corona as the error of the power law index is too large (S07; Section 3).
Thus, the current data are consistent with the conclusion that all BCGs with
$L_{\rm 1.4 GHz} > 10^{23}$ W Hz$^{-1}$ have a cool core, small or large.

BCGs with small coronae often host radio AGN as luminous as those BCGs in large cool cores.
We call these two classes the large-cool-core (LCC) class and the corona class. 
The cool cores in the LCC class are the cool cores that were generally referred
in the cluster papers. The corona class defined by this work also includes $\sim$ 5
systems with small cool cores (up to 25 kpc in radius) but larger than
typical coronae (less than 5 kpc in radius generally). We present the properties
of the X-ray sources associated with $L_{\rm 1.4 GHz} > 10^{24}$ W Hz$^{-1}$ radio AGN
in the corona class in Table 2 (also including non-BCGs). Some examples of BCG coronae
associated with strong radio AGN are also shown in Fig. 2 and 3.
We emphasize that the clusters or groups shown in Fig. 2 and 3 were considered
``noncool core'' systems by previous work (e.g., Mittal et al. 2009) so other
mechanisms (e.g., cluster merger) had to be used to explain strong radio AGN.
AWM4 is another example. It was once considered as a puzzle as it hosts a radio
AGN but lacks a large cool core from the \xmm\ data (Gastaldello et al. 2008).
The new \chandra\ data (released in late May, 2009, after the initial submission
of this paper) clearly show the presence of a small ($\sim$ 2 kpc radius),
thermal corona associated with the BCG (Fig. 3).

\subsection{The LCC class}

As shown in Fig. 1, the LCC class presents an intriguing correlation between the
cool core luminosity and the radio luminosity of the BCG, unlike the corona class. 
More luminous cool cores generally host more luminous radio sources, although the
scatter is large. From the BCES Orthogonal fit (Akritas \& Bershady 1996),
$L_{\rm 1.4 GHz} \propto L_{\rm X}^{1.91\pm0.20}$. Why does this correlation exist
for the LCC class? It is not clear that this is related to radio feedback.
On the other hand, environmental radio boosting can tilt
the LCC class to the observed trend, as more luminous cool cores are generally
bigger and are more capable to confine the radio lobes to stop adiabatic expansion
(e.g., Barthel \& Arnaud 1996; Parma et al. 2007).
Radio lobes can then be brighter and exist for a longer life time.
Thus, it is necessary to re-produce Fig. 1 for the radio nuclei, jets and lobes
separately. However, the relevant data are not available for most
radio AGN in this sample. We leave the question of the tilted LCC class to
future work as the corona class is the focus of this work.

The most intriguing result in Fig. 1 is for galaxy groups.
There are 19 groups with a cool core that is more luminous
than $6\times10^{41}$ ergs s$^{-1}$, but none of them hosts
a $L_{\rm 1.4 GHz} > 10^{24}$ W Hz$^{-1}$ AGN. Even at a lower radio
luminosity threshold (3$\times10^{23}$ W Hz$^{-1}$), groups in the corona
class outnumber groups in the LCC class by 18 to 1, while there are in
fact less cluster BCGs in the corona class than the LCC class (16 versus 27).
This deficiency of high $L_{\rm 1.4 GHz}$ groups in the upper portion
of the LCC class is not observed in clusters.
Two local systems ignored in our sample (but included in the B55 and the
HIFLUGCS-E samples, see Section 7.4), the Virgo cluster and the Fornax
cluster, also fit into this picture. The Virgo cluster hosts a
$L_{\rm 1.4 GHz} = 5.3\times10^{24}$ W Hz$^{-1}$ AGN in the center but
its temperature is $\sim$ 2.3 keV (e.g., Shibata et al. 2001). The central
radio source in the Fornax cluster is faint ($L_{\rm 1.4 GHz} = 3.2\times10^{22}$ W Hz$^{-1}$).
Is the deficiency of luminous group cool cores with a strong radio AGN
a selection bias from the heterogeneous nature of our sample?
Nearby groups are observed by \chandra\ either because they are bright
or they host bright central radio galaxies (so selected by the AGN panels).
Groups are then likely under-represented in the lower portion of the corona class,
but they are much less likely to be missed in the upper portion of the LCC class.
The absence of $L_{\rm 0.5-2 keV} > 6\times10^{41}$ ergs s$^{-1}$ cool cores
with a strong radio AGN for groups and poor clusters is also the reason why
there is a gap between two classes at high radio luminosity.
Nevertheless, this result needs to be examined with samples unbiased to the above
conclusion (e.g., optically selected).
If this result holds, it may point to a bigger role of radio outbursts
for group gas than for cluster gas, as the gas cores of groups are more vulnerable
to powerful radio outbursts compared to the larger, hotter gas cores of clusters.

\subsection{The corona class}

The properties of embedded coronae have been presented and discussed
in detail by S07. Jeltema et al. (2008) presented a similar analysis on
coronae in groups. We have added many more coronae for BCGs.
As shown in Fig. 1, X-ray luminosities of coronae show little correlation
with the radio luminosities of the galaxy, although coronae with strong
radio AGN are usually luminous. As emphasized in S07 and early in this paper,
small coronae in the upper portion of the corona class will be destroyed even
if only $<$ 1\% of AGN heating is acted inside the corona. Being in rich
environments (especially if in hot clusters), it is difficult to rebuild
these mini-cool cores from stellar mass loss, once they are destroyed.
S07 and this work show that very few coronae of massive galaxies should have been
destroyed in this way. Powerful radio jets may simply penetrate the coronal
atmosphere with very little energy deposition. However, if radio outbursts
have little impact inside a small corona, is radio heating still responsible
for quenching cooling inside a corona and turning off the nuclear activity?
This is the same cooling/heating question in large cool cores. Compared small
coronae with large cool cores, the global cooling is much weaker (but still
strong in the center) while heating is not apparently different from that in
large cool cores (Fig. 1). We will return to this issue in Section 7.

We also present the results for 22 non-BCGs with $L_{\rm 1.4 GHz} > L_{\rm 1.4 GHz, cut}$
(Fig. 4). Most non-BCGs do not have a confirmed corona but the upper limits
are usually high. Section 6 presents the faintest embedded corona known
(NGC~4709), which is much fainter than all upper limits in this work.
We also present the $L_{\rm Ks} - L_{\rm 0.5-2 keV}$ plot for all galaxies
in this work (Fig. 5). The corona class falls around the average relation
derived by S07, while the cool cores in the LCC class are much X-ray brighter.
We also included the expected emission from cataclysmic variables and
coronally active stars (Revnivtsev et al. 2008), which is always small.
As discussed in Section 6, the real contribution is even much smaller as
we always used the immediate local background that also includes much
stellar emission.
S07 showed that massive galaxies generally have luminous coronae. There
are some non-detections of coronae for massive galaxies in Fig. 5.
Generally the upper limits are still high. Interestingly, above $L_{\rm Ks}$
of 3.5 $L_{\rm Ks, *}$ (or $10^{11.61} L_{\rm Ks, \odot}$, the dotted line in Fig. 5),  
the galaxies associated with $L_{\rm 1.4 GHz} > 10^{23}$ W Hz$^{-1}$ AGN
are generally more X-ray luminous than those with weak radio AGN.
Massive BCGs with a moderate corona or no corona usually only have
weak radio AGN.
Thus, the strong connection between BCGs with strong radio AGN and
X-ray cool cores is not all because of their strong correlations with the
galaxy mass.

\subsection{X-ray AGN vs. Radio AGN}

We also examined the fraction of X-ray AGN ($L_{\rm 0.5 - 10 keV} > 10^{42}$
ergs s$^{-1}$) in our sample. Fainter AGN exist in many cases (Table 2). However,
the confirmation and determination of their properties are often affected 
by bright cluster cool cores, possible intrinsic absorption and LMXBs.
Moreover, fainter X-ray AGN cannot be well studied at
$z>$0.5 anyway. The results are summarized in Table 3. There are only
12 X-ray AGN above the threshold in total: NGC~1275, Cygnus A, Hydra A, PKS~2152-69,
five non-BCGs in A3667, A514, A754, A2142 and A3880, 3C264 in A1367, NGC~2484 and NGC~6251 
(Table 2). The fractions in all four sub-samples of Table 3 are small, although it indeeds
increases in the radio AGN sample.
We also notice that all BCGs with an $L_{\rm 0.5-2 keV} > 10^{42}$ ergs s$^{-1}$
AGN host an $L_{\rm 1.4GHz} > 10^{24}$ W Hz$^{-1}$ radio AGN.
Conversely, if a BCG hosts a radio source that is weaker than
$10^{24}$ W Hz$^{-1}$ at the 1.4 GHz, its X-ray AGN is never more luminous
than 10$^{42}$ ergs s$^{-1}$ in this sample (161 BCGs).

\section{A detailed case study: \ga\ in A3627}

The last section has depicted the general properties of the
corona class. How does a BCG corona associated with a luminous radio AGN
look in detail? In this section we present a detailed analysis of
a BCG corona with a strong radio AGN, \ga\ in A3627.
\ga\ is the brightest (not the most luminous) embedded corona in $kT >$ 3 keV
clusters (S07), because of its proximity.
A new 57.3 ks ACIS-S observation collects about 3800 source counts in the
0.5 - 3 keV band ($\sim$ 96\% from the gas). Before this observation,
NGC~3842 in A1367 (Sun et al. 2005a) had the most \chandra\ counts ($\sim$ 1000 from
a 43.2 ks ACIS-S observation) for an embedded corona.
Other coronae discussed in the literature (e.g., Coma, Vikhlinin et al.
2001; NGC1265 in Perseus, Sun et al. 2005b) only have $\sim$ 500 counts typically.
In fact, even including groups and poor clusters, only IC~4296 and NGC~315
have brighter coronae. However, both are in poor groups so the surrounding
ICM has a comparable temperature. Both also host bright nuclear sources
and there is a bright X-ray jet in NGC~315 (Worrall et al. 2003), which causes
some trouble for the analysis of the X-ray gas. Radio AGN associated with these
two galaxies are $>$ 17 times fainter than \ga's. Thus, \ga\ is the best case
of a luminous corona associated with a strong radio AGN for a detailed analysis.
In S07, we present the results from an 14.1 ks ACIS-I observation targeted
on a cluster position that is 5$'$ from \ga. Here we present the results from a
much deeper targeted ACIS-S observation in cycle 8 (PI: Sun).

At $z=0.01625$ (Woudt et al. 2007), A3627 ($kT \sim$ 6 keV) is the closest massive cluster.
\ga\ is the BCG of A3627. ESO~137-008 (also
with a small corona) has the same $K_{\rm s}$ band magnitude as \ga's, but its velocity
dispersion is much smaller (see Table 2 of S07). The radio WAT source associated with \ga,
PKS~1610-60 (Jones \& McAdam 1996), is one of the brightest radio sources in
the southern hemisphere and its 1.4 GHz luminosity is 53\% higher than
NGC~1275's (Fig. 1; Table 2). The \xmm\ mosaic image of A3627 is shown in Fig. 6,
along with the radio contours from SUMSS. One can observe evidence of the
interaction between the radio lobes and the ICM in Fig. 6, which implies that
\ga\ is not far from the cluster gas core. For the assumed cosmology (Section 1), the angular
scale is 0.327 kpc / arcsec and the luminosity distance of A3627 is 69.6 Mpc.

\subsection{\chandra\ data analysis}

The observation of \ga\ was performed with ACIS-S on 2007 July 8. Standard \chandra\
data analysis was performed (see Section 3). No background flares were present.
The effective exposure is 57.3 ks for the S3 chip. The observation was performed
in the VFAINT mode. However, we only use the FAINT mode for the analysis
of the corona source, as the VFAINT filtering causes a 2\% loss on
the source counts, mostly at the center of the corona. We still
use the VFAINT mode for the analysis the surrounding ICM.
In the analysis of the corona, we also removed the pixel randomization.
We used an absorption column density of 1.73$\times10^{21}$ cm$^{-2}$ from the
Leiden/Argentine/Bonn HI survey (Kalberla et al. 2005). If we choose to fit
the absorption column from the \chandra\ spectra, the best fits are always
consistent with the above value. This absorption column is lower than
the previous value from Dickey \& Lockman (1990), 2.0$\times10^{21}$ cm$^{-2}$,
which had been used in previous work (e.g., B$\ddot{\rm o}$hringer et al. 1996; S07).

\subsection{The properties of the surrounding ICM}

Before we study \ga's corona, the properties of its surrounding ICM were examined
to constrain the ambient gas temperature and pressure. As discussed in the Appendix
of S07, the soft X-ray background is high in the direction of A3627. S09
developed a method to constrain the local X-ray background with the stowed
background. Following the S09 method, we derived a local soft X-ray background
flux surface density of (12.1$\pm$1.0) $\times10^{-12}$ ergs s$^{-1}$ cm$^{-2}$ deg$^{-2}$
in the 0.47 - 1.21 keV band. The RASS R45 flux measured from 1 - 2 deg annulus
centered on A3627, excluding a bright PS, is 290 $\times 10^{-6}$ cnts s$^{-1}$ arcmin$^{-2}$.
These two fluxes well match their average relation shown in the Fig. 2 of S09.
The hotter thermal component of the local X-ray background has a temperature of 0.33$\pm$0.03 keV
(see the local background model in S09), which is consistent with the
trend that its temperature increases to $\geq$ 0.3 keV in the high R45 flux regions (S09).

With the local X-ray background determined, the temperature profile of the
surrounding ICM was derived ($\sim$ 7 keV, Fig. 7). The ICM abundance from the joint fit of the
three radial bins is 0.32$\pm$0.09 solar, which is much lower than that of the small corona
(the next section).
The ICM electron density from the \rosat\ data is $\sim 1.9\times10^{-3}$ cm$^{-3}$
at the projected position of \ga\ (B$\ddot{\rm o}$hringer et al. 1996).

\subsection{The properties of \ga's corona}

The 0.5 - 5 keV \chandra\ count image of \ga\ is shown in Fig. 6. Beyond the bright
core, the corona is more extended towards the south, which implies \ga's motion to
the north. This is consistent with the small bending of its radio lobes (Fig. 6).
To better show the central part of the corona, we applied the Subpixel Event
Repositioning tool developed by Li et al. (2004) to the data to make the 1/4
subpixel image (the right bottom panel of Fig. 6). Within the central 0.5 kpc,
the gas is flattened along the north-south direction, which is also perpendicular
to the direction of the radio jets.

We derived the profiles of the X-ray surface brightness, temperature (projected and
deprojected), density, entropy, cooling time and pressure for the corona, as shown
in Fig. 7. In the spectral analysis, the local background is extracted from the
34$''$ - 120$''$ (11.1 - 39.2 kpc) annulus. VAPEC was used in each annulus, as
it fits the spectral lines better. We used the classical ``onion-peeling'' method
for deprojection. For such a massive galaxy, the LMXB emission should be significant.
We used the $L_{\rm X, LMXB} - L_{\rm Ks}$ scaling relation derived by Kim \&
Fabbiano (2004). As there are no \hst\ data of \ga, we simply used the \hst\ stellar
light profile of NGC~3842 for the stellar light profile of \ga\ (Sun et al. 2005a),
as two BCGs have similar 2MASS $K_{\rm s}$ luminosities (0.22 mag difference). After
adjusting to \ga's $K_{\rm s}$-band luminosity and A3627's local background, one
can see that the LMXB light + a flat local background well describes the observed
profile beyond $\sim$ 4.2 kpc radius. In fact, the net spectrum in the 13$''$ - 34$''$
annulus (with the 34$''$ - 120$''$ spectrum as the local background) can be well
fitted by a power law with a photon index of 1.7. Its flux is also consistent with
the expected LMXB flux in the region (Table 4). Thus, we conclude that the corona is
pressure confined at $\sim$ 4.2 kpc radius.
In fact, the electron pressure ratio across the 4.2 kpc boundary is 1.16$\pm$0.51,
consistent with pressure equilibrium. Obviously, there are large jumps of entropy
and cooling time across the boundary. Across the coronal boundary, heat conduction has to be
suppressed by a factor of at least $\sim$ 230 from the Spitzer value, using the
ratio of the heat conduction flux and the total X-ray flux (the method used in
S07). Without any suppression, this small corona will be evaporated in
$\sim$ 8 Myr (S07).
Inside the boundary, the corona has low entropy and short cooling time (Fig. 7).
that is typical for the center of large cool cores.
The properties of \ga's corona are summarized in Table 4.

As the abundances in each annulus are all consistent with the same value (from the
VAPEC model), we fix them together.
The best-fit abundances from the VAPEC model are listed in Table 4.
We also derived the abundance ratios: Si/Fe = 1.47$^{+0.94}_{-0.58}$,
Fe/O = 2.17$^{+3.60}_{-1.42}$, and Fe/Mg = 0.50$^{+0.35}_{-0.20}$.
These ratios can be compared with the theoretical estimates from
Iwamoto et al. (1999): Fe/O = 0.26, Si/Fe = 3.03, and Fe/Mg = 0.26
for SNII, Fe/O = 27-75, Si/Fe = 0.54-0.62, and Fe/Mg = 31-69 for SNIa. 
Thus, the abundances of the corona is not only enriched by SNIa.

As shown in Fig. 7, the central density is high. Can we constrain the gas
properties around the Bondi radius? The mass of the central SMBH can be estimated
from the $K_{\rm s}$-band luminosity of the galaxy (Marconi \& Hunt 2003, see Section 7.1),
1.61$\times10^{9}$ M$_{\odot}$. The Bondi radius is then 62 pc (or 0.19$''$) for
a central temperature of $\sim$ 0.8 keV (Fig. 7). This scale is unresolved.
However, we can still constrain the range of gas density and entropy around
the Bondi radius, from the measured total X-ray emission within the innermost
bin. We assume $n_{\rm e} = n_{\rm e,0} (r/r_{0})^{-\alpha}$, where $r_{0}$ is the
Bondi radius and $n_{\rm e,0}$ is the electron density at $r_{0}$.
Of course, $\alpha <$ 1.5. We integrated the density model to compare with the
observed X-ray luminosity within the innermost bin, assuming a constant
emissivity. As shown in Fig. 8, the gas density at the Bondi
radius is always $\leq$ 1.8 cm$^{-3}$, even when the PSF correction is
considered. Similarly to compare the model with the emission-weight temperatures
observed, we find that the gas temperature at the Bondi radius is 0.8 - 0.9 keV.
These results will be used in Section 7.1 to examine whether Bondi accretion
is sufficient to power the radio AGN of \ga.

In spite of its high central density, the size of \ga's corona is small so the total gas
mass is very small, (1.78$\pm$0.19) $\times 10^{8}$ M$_{\odot}$ within 13$''$ (or 4.24 kpc) radius.
This can be compared with the typical gas mass within 10 Gyr cooling radius
for luminous cool cores in groups and clusters, $\sim 10^{11} - 5\times10^{12}$ M$_{\odot}$
(S09; Cavagnolo et al. 2009). Owing to the vast contrast, S07 named the embedded
coronas as mini-cooling cores. As we will discuss in Section 7, this kind of mini-cooling
cores is sufficient to fuel powerful FR-I radio galaxies. As shown in Fig. 1 and Table 2,
the X-ray luminosity of \ga's corona is not particularly high in the corona class.
Thus, we expect that other luminous BCG coronae would have
similar properties to those of \ga's.

\section{How faint an embedded corona can be?}

There are upper limits of the coronal emission in Fig. 1, 4 and 5, so it is interesting
to know how faint an embedded corona can be. Strong ram pressure in groups and clusters
generally only allows coronae with dense cores to survive (S07). While many BCGs have
luminous coronae ($L_{\rm 0.5-2 keV} \sim 10^{41}$ ergs s$^{-1}$, Fig. 4 and 5),
faint embedded coronae with moderate cool cores do exist. For BCGs associated
with luminous radio AGN, the faintest corona known is the one associated with
NGC~3862 (or 3C~264) in A1367 ($L_{\rm 0.5-2 keV} \sim 1.4 \times 10^{40}$ ergs s$^{-1}$,
Table 2; S07). One can see that its luminosity is comparable to or smaller than all upper limits
for non-detections in this work (Fig. 4 and 5). How faint can a BCG corona be?
In this section we discuss the corona of NGC~4709, which has the lowest X-ray
luminosity known for an embedded corona in a galaxy more luminous than $L_{*}$.

NGC~4709 is the dominant galaxy of a shock-heated subcluster falling into the
Centaurus cluster (Churazov et al. 1999). A faint corona was detected near the edge
of the S1 chip in a 34.3 ks \chandra\ observation (S07). Here we present the results
from our new targeted observation. We adopt a distance of 35.3 Mpc for NGC~4709
(Tonry et al. 2001), which is much smaller than the distance we used in S07 (49.4 Mpc).
The angular scale is 0.167 kpc / arcsec. The observation of NGC~4709 was performed
with ACIS-S on 2007 March 25. No background flares were present.
The effective exposure is 29.7 ks. An extended but faint source is detected at
the position of NGC~4709. A strong iron L-shell hump centered at $\sim$ 0.9 keV
clearly shows the presence of the thermal gas. The new targeted observation allows
better removal of the local background. Half of the 0.5 - 2 keV counts are from
a hard power-law component (likely LMXBs) and about 90 counts are from the emission of
the thermal gas. The gas temperature is 0.78$\pm$0.09 keV with a
fixed abundance of 0.8 solar (see S07). The temperature difference from S07 comes
from the different contribution of the hard power-law component in the spectral
fits. The S07 fit to the old S1 data allows little flux from the hard component.
The rest-frame 0.5 - 2 keV luminosity is 1.8$\times10^{39}$ ergs s$^{-1}$, which
makes it the faintest corona known in clusters (S07 and this work). We compare its \chandra\
surface brightness profile with that of \ga\ in Fig. 9. The contribution of the
LMXB light has been removed, assuming that it follows the stellar light. One
can see the vast difference. The corona boundary is $\sim$ 2.0 kpc.
Assuming a constant emissivity for the corona, the central electron density
of the gas is 0.034 cm$^{-3}$ and the total gas mass is 2.9$\times10^{6}$ M$_{\odot}$
within a 2 kpc radius (Table 4). The central cooling time is $\sim$ 0.13 Gyr
so it is still a moderate cool core.
The properties of NGC~4709's corona are summarized in Table 4 to compare with
those of \ga's corona.

We indeed notice that faint thermal
emission can also come from the integrated emission of cataclysmic variables and
coronally active stars (e.g., Revnivtsev et al. 2008).
This emission component is linearly scaled with
galaxy's $K_{\rm s}$-band luminosity (Revnivtsev et al. 2008). As shown in Fig. 5,
the expected luminosity of this component gets close to that observed for NGC~4709,
although it is far smaller than those of luminous coronae.
However, one needs to remember that in our analysis of small coronae,
local background is always used. For NGC~4709, the global spectrum is
extracted from an aperture of 2.23 kpc in radius, while the local background
is from the 2.23 kpc - 5.82 kpc annulus. We analyzed the 2MASS $K_{\rm s}$ band
image and found that the net stellar emission within 2.23 kpc radius is only
$\sim$ 3.5\% of galaxy's total light, after subtracting the local background
from the 2.23 kpc - 5.82 kpc annulus. Moreover, NGC~4709's X-ray source is
more peaked than the optical light. Thus, we are confident that NGC~4709 has
an X-ray gaseous halo.

Coronae like NGC~4709's are so faint that they are easily overlooked at $z>$0.03.
Coronae like 3C264's can also be easily overlooked at $z>$0.07, especially when
a bright X-ray nuclear source is present (the case for 3C264).
This fact needs to be kept in mind when $z \geq$0.05 FR-I radio galaxies or cluster BCGs
are studied. A significant X-ray gas component that is sufficient to fuel the central
AGN may elude detection easily.

\section{Discussion}

\subsection{Fueling radio AGN}

What fuels the radio AGN in groups and clusters? We focus on radio AGN associated
with a small corona, as radio AGN in the LCC class have been widely discussed in
the literature. Even for small coronae,
there is sufficient amount of gas cooled from the hot phase to fuel the radio AGN.
The maximum mass deposition rate from cooling (assuming steady and isobaric) is:
$\dot{M}_{\rm cooling} \approx 2\mu$$m_{\rm p}L_{\rm X, bol}/5kT = 0.44$
($L_{\rm bol}/10^{41}$ ergs s$^{-1}$) ($kT$/0.9 keV)$^{-1}$ M$_{\odot}$/yr.
We calculate $\dot{M}_{\rm cooling}$ for each corona in the upper portion of
the corona class (Fig. 10).
The bolometric correction is made case by case. For upper limits, we assume a
temperature of 0.7 keV (S07). The required SMBH accretion rate is estimated from the
$L_{\rm 1.4 GHz}$ - jet power relation by B\^{i}rzan et al. (2008), assuming a mass-energy
conversion efficiency of 0.1. As shown in Fig. 10, the required SMBH accretion rate
is always small and there should be enough amount of cooled gas in coronae.
The reality is however more complicated. First, the above cooling rate is reduced
if there are heat sources. Even if strong radio outbursts deposit very little heat
inside a corona, SN heating can be significant as discussed in S07.
On the other hand, the stellar mass loss rate from evolved stars is
significant inside coronae, 0.2 - 0.8 M$_{\odot}$/yr (Faber \& Gallagher 1976;
S07). However, both the SN heating and the stellar mass loss should follow the
stellar light profile, which is much shallower than the X-ray emission profile
of coronae (S07).
The detailed energy balance and evolution of gas in different phases
is unclear. However, as argued in S07, cooling in the central kpc of a luminous
corona (where most X-ray emission comes from) should overwhelm heating so
the actual mass deposition rate should not be reduced much from the above estimates.
Second, the $L_{\rm 1.4 GHz}$ - jet power relation in B\^{i}rzan et al. (2008)
was derived from radio AGN in the LCC class, where the radio luminosity of lobes
may be enhanced by the high ICM pressure of large cool cores (e.g., Barthel \& Arnaud 1996).
Moreover, the B\^{i}rzan et al. (2008) relation likely underestimates the jet power
by missing power from undetected weak shocks and cavities. These two factors
should not increase the jet power by more than a factor of a few
(McNamara \& Nulsen 2007) so cooling of small coronae should provide enough
fuel for their radio AGN.

Are radio AGN fueled by hot gas (e.g., Best et a. 2005; Croton et al. 2006;
Allen et al. 2006; Hardcastle et al. 2007)? Typically for hot accretion,
Bondi accretion is assumed (e.g., Hardcastle et al. 2007). The Bondi accretion rate is:
$\dot{M}_{\rm Bondi} = 0.0042 (K / 2$ keV cm$^{2})^{-3/2} (M_{\rm BH} / 10^{9} M_{\odot})^{2} M_{\odot}$/yr.
$K$ is the gas entropy at the Bondi radius. The Bondi radius in this sample
(10 - 150 pc) is always unresolved. However, as shown in Section 5.3, the gas
entropy can still be constrained from the observed X-ray luminosity of the
innermost bin. For the best-studied case, \ga, we take the gas entropy at the Bondi
radius as 0.8$\pm$0.3 keV cm$^{2}$ (Section 5.3).
There are four luminous coronae that we performed detailed studies before,
NGC~3842 and NGC~3837 (Sun et al. 2005a), NGC~1265 (Sun et al. 2005b) and
3C~465 (S07). Their central entropy values are $\leq$ 1.1-1.8 keV cm$^{2}$.
For the faintest corona of a massive galaxy, NGC~4709's corona, the central
entropy is $\sim$ 6.5 keV cm$^{2}$ (Section 6). As we cannot perform a similar
analysis for \ga's corona on other coronae, we simply assume
an average entropy of 1.5 keV cm$^{2}$, with an uncertainty range of 0.5 - 3 keV cm$^{2}$
(note that Hardcastle et al. 2007 used an entropy value of 1.11 keV cm$^{2}$).
The SMBH mass is estimated from the 2MASS $K_{\rm s}$ band luminosity of the galaxy
(Marconi \& Hunt 2003),
log($M_{\rm SMBH} / M_{\odot}) = 9.47 + 1.13$ log($L_{\rm Ks} / 10^{12} L_{\odot}$).
The results are also shown in Fig. 10.
A large Bondi accretion rate needs a massive black hole and low-entropy
surrounding gas. Both requirements point to massive galaxies (recall the
$L_{\rm X} - L_{\rm Ks}$ correlation for coronae, e.g., S07). Thus, low-mass
galaxies with a faint corona may not power their radio AGN through hot
accretion (Fig. 10), especially if the required SMBH accretion rate is higher
for the reasons presented in the last paragraph.
However, it should be aware that both the Marconi \& Hunt (2003) relation
and the B\^{i}rzan et al. (2008) relation have large scatters ($\sim$ 0.5 - 1 dex),
which makes it impossible to reject hot accretion for any sources in the plot.
On the other hand, SMBH spin can reduce the required accretion rate
(e.g., Wilson \& Colbert 1995; McNamara et al. 2009).

Besides hot gas, dust and molecular gas are also present in early-type galaxies
and BCGs. Dust with mass $\leq 10^{4} - 10^{5}$ M$_{\odot}$ was detected in
$\sim$ 80\% of nearby large early-type galaxies (e.g., van Dokkum \& Franx 1995).
The origin of dust includes mergers with dust-rich dwarfs and a stellar origin
(Mathews \& Brighenti 2003).
Interestingly, the dust detection rate is higher in radio-jet galaxies than in
non radio-jet galaxies (e.g., de Koff et al. 2000; Verdoes Kleijn \& de Zeeuw 2005;
Sim\~{o}es Lopes et al. 2007). For galaxies with dust detections, FR-I galaxies
also have more dust than radio-quiet ellipticals.
The dust in FR-I galaxies is generally situated in sharply defined disks on small
($<$ 2.5 kpc) scales, while radio jets are generally perpendicular to the dust disk
(e.g., de Koff et al. 2000). Sometime the dust disk is warped (e.g., 3C~449 in our
sample, Tremblay et al. 2006), likely in the process of settling down to be regular
disks or are being perturbed (Verdoes Kleijn \& de Zeeuw 2005). The dust-jet
connection is still not well understood (e.g., which impacts which?) but dust may
play a role to fuel the central SMBH.
MIR emission from PAHs and dust in some FR-I galaxies was also found from the
\spi\ data (e.g., Leipski et al. 2009).

There is also mounting evidence of CO detections in radio galaxies (e.g.,
Lim et al. 2000; Evans et al. 2005; Prandoni et al. 2007). Three galaxies in our
sample belongs to this growing class, 3C~31 (a molecular gas mass of 4.8-10$\times10^{8}$ M$_{\odot}$,
Lim et al. 2003; Evans et al. 2005; Okuda et al. 2005), 3C~264 or NGC~3862 (a molecular
gas mass of 2.6$\times10^{8}$ M$_{\odot}$, Lim et al. 2003) and 3C~449 (a molecular gas mass of
$2.4\times10^{8}$ M$_{\odot}$, Lim et al. 2003).
The CO emission of 3C~31 and 3C~264 exhibits a double-horned line profile characteristic
of a rapidly-rotating disk (Lim et al. 2000; Okuda et al. 2005). At least eleven other galaxies in our
sample were undetected in these CO observations, with upper limits of
(1-5)$\times10^{8}$ M$_{\odot}$ (all mass values have been adjusted to our cosmology).
It is intriguing that the estimated mass values or upper limits of the molecular gas are comparable
or even larger than the mass values of the X-ray gas of their coronae, although the cold gas
can in principle be produced through cooling over $<$ 1 Gyr.
It should also be aware that CO detections are also present in the general samples of
BCGs in large cool cores (Edge 2001; Edge \& Frayer 2003; Salom${\rm \acute{e}}$ \& Combes 2003).
Some of BCGs with significant CO detections only have weak radio AGN (e.g., A262, 2A~0335+096
and A2657 with $L_{\rm 1.4 GHz} = 0.09-1.0\times10^{23}$ W Hz$^{-1}$).
More detailed molecular data of BCGs with and without strong radio AGN will better
reveal the connection between the radio activity and the cold gas component.
A related question is on the nuclear star formation in small coronae.
Rafferty et al. (2008) and Cavagnolo et al. (2008) showed that star
formation turns on when the central cooling time of the gas falls below $\sim$ 0.5 Gyr
or the gas entropy falls below $\sim 30$ keV cm$^{2}$ (at $r$ = 3-4 kpc for \ga, Section 5).
Will this be the case for small coronae?
On the other hand, the energy balance and transfer between gas in different
phases (molecules, dust, stellar winds and 10$^{7}$-K X-ray gas) is an interesting problem
to explore.

To sum up, cooling of the X-ray coronae can provide enough fuel to the central SMBH,
provided that the cooled materials can reach the very center. Bondi accretion may be
sufficient for the massive black holes in a luminous X-ray corona, but likely not the
only answer for the strong radio AGN in the corona class. Fig. 10 also does not necessarily
argues for hot accretion for massive galaxies as there are many uncertainties. Besides
ones mentioned above, it should be aware that a small amount of angular momentum of
the hot gas can largely reduce the accretion rate (e.g., Proga \& Begelman 2003).
The large scatter of the $L_{\rm 1.4 GHz}$ - jet power relation also implies that
some strong radio AGN will require much higher mass accretion rates than the
average values.
On the other hand, dust and molecular gas co-exist with the 10$^{7}$-K X-ray gas in
radio AGN (including some in our sample), which may bring cold accretion into play.
One issue we did not discuss is galaxy merger as many BCGs are dumbbells.
Although the connection between the various gas components is unclear, we suggest that
the existence of an X-ray corona with high pressure can effectively shield dust and cold gas from strong
evaporation and stripping by the ICM, especially in hot clusters.

\subsection{Are coronae decoupled from the radio feedback cycle?}

If small coronae are responsible to fuel strong radio AGN (either
through cold accretion of the cooled materials or through hot accretion),
is radio heating responsible to offset strong cooling inside coronae?
As a complete radio feedback cycle is generally assumed in large
cool cores (summarized in e.g., McNamara \& Nulsen 2007), another
way to put the question is: is the radio feedback cycle complete in
small coronae as in large cool cores?
The ICM surrounding a corona is clearly decoupled from the feedback cycle,
simply absorbing heat from unrelated radio outbursts. 
As discussed in S07 and early in this paper, it requires a fine
tuning for a strong radio outburst to offset cooling inside a small corona
without completely destroying it, which is not a problem for large cool cores. 
This is more a problem when it is considered that $\sim$ 20\% of BCGs
have $L_{\rm 1.4 GHz} > 10^{24}$ W Hz$^{-1}$ AGN (so the active period of
strong heating is a large portion of galaxy's life time, Lin \& Mohr 2007) and the
ubiquity of small coronae associated with massive galaxies (S07).
Other heat sources inside coronae do exist as discussed
in S07, e.g., stellar mass loss and SN heating. However, these heating
terms should follow the shallow stellar light profile so cooling
will always overwhelm within 2 - 3 kpc from the nucleus (S07).

Although strong radio outbursts may simply penetrate small coronae,
weak and more frequent radio outbursts are able to release significant
amount of heating within the central a few kpc to offset cooling.
Nearby examples include M84 (Finoguenov \& Jones 2001), NGC~4636
(Jones et al. 2002) and NGC~4552 (Machacek et al. 2006) in the distance of
the Virgo cluster. Although their radio AGN are weak ($L_{\rm 1.4 GHz}$ =
0.024 - 1.89 $\times10^{23}$ W Hz$^{-1}$), their coronae are significantly disturbed.
The radio outbursts in these systems are weak, with energy from $\sim 1.4\times10^{55}$ ergs
to $\sim 6\times10^{56}$ ergs (Jones et al. 2002;
Machacek et al. 2006; Finoguenov et al. 2008), which can be compared with the biggest
radio outburst known in clusters, $\sim 1.2\times10^{62}$ ergs in MS~0735.6+7421
(McNamara et al. 2005, 2009). A $10^{56} - 10^{57}$ ergs outburst will not
disrupt a luminous corona too much (e.g., 4$\int P dV \sim 3\times10^{57}$ ergs
for \ga's corona), which is about the amount of gentle heating required to offset cooling
in a small corona.
Thus, gentle heating from weak radio AGN may complete the feedback cycle in
coronae. However, this is not exactly the answer to strong radio AGN in the
upper portion of the corona class, as it is unknown what will turn them to
a lower activity state if the strong radio heating is not involved in the
feedback cycle. Maybe radio heating of the coronal gas is only important in
weak outbursts, while heating is only on the surrounding ICM in strong outbursts.
During the period of strong outbursts, SN heating inside coronae can still
offset some cooling and the stellar mass loss can compensate a significant portion
of gas cooled out of the hot phase (S07).
The detail of energy balance and transfer in these embedded mini-cool cores
is beyond the scope of this work. It would also help to know the difference of
the radio AGN populations associated with a small corona and associated with a large cool core,
e.g., duty cycles. The current large sample studies (e.g., Best et al. 2007;
Lin \& Mohr et al. 2007) lack enough X-ray data to divide two classes. 

A related question is on the origin of the gap in Fig. 1, or the absence of $\sim 10^{42}$
ergs s$^{-1}$ cool cores at $L_{\rm 1.4 GHz} > 10^{24}$ W Hz$^{-1}$. The gap is not
observed at low $L_{\rm 1.4 GHz}$. Systems that should fill the gap are mainly groups.
Why are there so few luminous group cool cores with strong radio AGN? Have that kind
of group cool cores been transferred into the corona class in a powerful radio outburst?
We estimate the required outburst energy to heat group cool cores beyond $\sim$ 5 kpc
radius, from the group pressure profiles derived by S09. Outburst energy of
$10^{59} - 10^{60}$ ergs (4$\int P dV$) is required. The corresponding power in 10$^{8}$ yr
is 3$\times10^{43} - 3\times10^{44}$ ergs s$^{-1}$. A radio AGN with $L_{\rm 1.4 GHz}$ of
$\sim 10^{24}$ W Hz$^{-1}$ can provide that amount of energy on average (B\^{i}rzan et al. 2008),
especially if the biasing factors to the B\^{i}rzan et al. relation discussed in the
last section are considered.

\subsection{The implications of the corona class}

The existence of a significant number of strong radio AGN in the corona
class has important implications for the radio AGN heating and formation
of large cool cores. As shown in this work and some previous work
(e.g., Hardcastle et al. 2007), ISM accretion in early-type galaxies is sufficient to
power FR-I radio AGN, long before a large cluster cool core is formed.
The radio outburst injects a large amount of heat
into the surrounding ICM and is capable to 
destroy embryonic large ICM cool cores beyond the central several kpc.
Therefore, this provides another way to prevent formation of
large cool cores in some clusters, besides the scenario of major
mergers at an early stage proposed by Burns et al. (2008).
Nevertheless, large cool cores do form. Radio AGN heating in massive
clusters may not be strong enough to balance cooling (e.g., Best et al. 2007).
A dense corona of a BCG may also not form because of high SN rate
and merging rate so the corona feedback has never been triggered
to destroy an embryonic ICM cool core.
It would be useful to know the percentage of large cool cores (or coronae)
for BCGs as a function of the cluster mass and redshift. Our heterogeneous sample indeed implies
that the fraction of coronae for BCGs increases when the cluster mass decreases
(Fig. 1).
 
The dense coronal gas is also important to maintain the collimation of
radio jets (e.g., Fabian \& Rees 1995). There is no evidence for slower
milli-arcsec scale jets of FR-I sources in comparison to FR-II sources
(Giovannini et al. 2001) so the FR-I jets must slow down from the pc to kpc
scale. The existence of a dense corona provides extra pressure to decelerate jets.
In the case of \ga,
$\int\limits_{0}^{r_{\rm cut}}P_{\rm ISM} dr / (P_{\rm ICM} r_{\rm cut}) = 7.7$.
The best-fit $\beta$-model for the density profile was used (Fig. 7) and
$r_{\rm cut}$ = 4.24 kpc. This is only a comparison of pressure, without
considering the slowing of jets with time. Nevertheless, this simple comparison
shows that jet flaring would be at much larger radii if the radio AGN is
``naked'' in the ICM. While a corona helps to decelerate jets, its small
size also allows the bulk of the jet energy to be able to transfer to large
radii of the system. This was revealed in the detailed modeling for jet
deceleration in 3C~31 (Laing \& Bridle 2002). They concluded that a small
X-ray cool core, associated with the BCG NGC~383 rather than with the
surrounding ICM, is required for the jet to decelerate without disruption.
This is in contrast with large cool cores, where radio jets and lobes are
much more likely to be contained, with less energy being able to deposit
at the outskirts of the system. This effect is especially important
for groups, where the central cool cores are generally smaller than those
in rich clusters with a large cool core.

Hardcastle \& Sakelliou (2004) studied jet termination in WAT radio sources.
They presented an anti-correlation between the jet termination length and
the cluster temperature. One scenario listed in the paper is that the jet
disruption coincides with the ISM/ICM interface. Seven WATs in their sample
are in our sample. Clearly from this work and S07, the ISM coronae have radii of
typically a few kpc (also see the pressure argument in S07), at most 9-10 kpc for
3C~465 (S07). The coronae are much smaller than the derived jet termination
length (12-74 kpc) by Hardcastle \& Sakelliou (2004) so the above scenario can be ruled out.
On the other hand, as shown in Sun et al. (2005a; 2005b) and S07, radio
jets often turn on after transversing the corona/ICM boundary. 

The existence of a large number of BCGs in the corona class shows that the
X-ray cool cores of BCGs have a wide range of luminosities and masses. Many
systems that were considered as ``noncool core'' clusters (e.g, many
in the HIFLUGCS sample, Chen et al. 2007; Mittal et al. 2009) in fact have
BCG coronae. {\em Is the traditional cool core / noncool core dichotomy too simple?}
We suggest that a better alternative is to use the cool core distribution function,
with the enclosed X-ray luminosity or gas mass. This better describes the
X-ray gas component with short cooling time associated with BCGs. It
naturally explains the existence of strong radio AGN in the so-called
``noncool core'' clusters (e.g., some in Fig. 2 and 3).

\subsection{The B55 and the extended HIFLUGCS samples}

Our sample (Table 1) includes nearly all systems in the B55 and the extended
HIFLUGCS samples. The B55 sample (55 clusters, Peres et al. 1998) is a hard X-ray
selected sample that has been fully covered by \chandra. There are no groups
($kT<$ 2 keV) in the B55 sample. The basic HIFLUGCS sample has 63 groups and
clusters (Reiprich \& B$\ddot{\rm o}$hringer 2002) that has been fully covered by
\chandra. Reiprich \& B$\ddot{\rm o}$hringer (2002) also listed 11 systems that
are brighter than the HIFLUGCS flux limit but within 20 deg from the Galactic plane
so that they are not included in the HIFLUGCS sample. We include them and call
the resulting sample the extended HIFLUGCS (HIFLUGCS-E) sample. Only the
Antlia group does not have \chandra\ data, but its BCG is radio quiet anyway.
Instead of marking the B55 and the HIFLUGCS-E systems in Fig. 1, we plot the
cooling time at 10 kpc radius of the BCG versus the 1.4 GHz luminosity of
the BCG (Fig. 11). The cooling time profiles come from S09,
Cavagnolo et al. (2009) and our own work.
We choose a radius of 10 kpc to well separate small
coronae from large cool cores. Mittal et al. (2009) presented a similar
plot for the HIFLUGCS sample, but they used isobaric cooling time at
0.004 $r_{500}$ (2 - 6 kpc) and the total radio luminosity. They ignored
the presence of small coronae so other mechanisms (e.g., cluster merger)
were used to explain strong radio AGN in some ``noncool core'' clusters
that are all found to host small coronae in this work. Cavagnolo et al. (2008)
presented a plot of the central entropy versus the radio luminosity.
Their conclusion of the association between low entropy gas and the strong
radio AGN is still consistent with ours. In Fig. 11, we did not include A1367
and A754 as they are irregular clusters with the X-ray peak far from their
BCGs. In the plot for the HIFLUGCS-E sample, A2163, as the most distant system
($z$=0.203), was excluded.

As shown in Fig. 11, there is a general anti-correlation between the radio
activity and the central gas cooling time (e.g., at 10 kpc), especially if groups are excluded.
However, there are three outliers in the B55 sample and seven outliers in the
HIFLUGCS-E sample. All these BCGs have small coronae. More exactly, the BCGs
of A3532 and A3376 are only flagged as soft X-ray sources (or candidates of coronae,
see Section 3) as the observations are not deep enough (9.5 - 64 ks with ACIS-I)
at their redshifts ($z$=0.046-0.055). As emphasized in Section 3, we are confident
that most soft X-ray sources are genuine thermal coronae.
At $L_{\rm 1.4 GHz} > 10^{24}$ W Hz$^{-1}$, the fraction of coronae increases
when the cluster flux limit of the sample decreases, 3 out of 16 in the B55 sample
versus 7 out of 21 in the HIFLUGCS-E sample. The fraction is 30 out of 49 
(14 out of 33 for $kT >$ 2 keV systems) in our sample although ours is not a
flux-limit sample. This is not surprising as small flux-limited samples are
biased to the local luminous LCC clusters. 
Owing to its high flux limit, groups in the extended HIFLUGCS sample are luminous
groups with large cool cores and none of their BCGs has strong radio activity
($L_{\rm 1.4 GHz} > 10^{24}$ W Hz$^{-1}$), while these are many such groups in our sample (Fig. 1).

\section{Conclusions and further questions}

We present a systematic study to search for X-ray cool cores associated with
161 BCGs and 74 strong radio AGN ($L_{\rm 1.4 GHz} > 10^{24}$ W Hz$^{-1}$) in
152 nearby groups and clusters (186 galaxies in total), selected from the
\chandra\ archive, including nearly all systems in the B55 and the extended
HIFLUGCS samples. The main conclusions of this work are:

1) All 69 BCGs with strong radio AGN ($L_{\rm 1.4 GHz} > 2 \times 10^{23}$ W Hz$^{-1}$)
have X-ray cool cores with a central isochoric cooling time of $<$ 1 Gyr
(Section 4 and Fig. 1).
In fact, there are only two non-detections of cool cores out of 81 BCGs above
$L_{\rm 1.4GHz}$ of $10^{23}$ W Hz$^{-1}$. Upper limits of both non-detections are high
so we claim that every BCG with a $L_{\rm 1.4 GHz} > 10^{23}$ W Hz$^{-1}$
AGN has an X-ray cool core. This conclusion also holds in the B55 and the extended HIFLUGCS
samples (Section 7.4). The BCG cool cores can be divided
into two classes, large ($r_{\rm 4 Gyr} >$ 30 kpc) and luminous
($L_{\rm 0.5-2 keV} \geq 10^{42}$ ergs s$^{-1}$) cool cores like Perseus's
cool core, or small ($\leq$ 4 kpc in radius typically) coronae like \ga's in A3627 (Section 5,
Fig. 6-7 and S07, also see Fig. 2-3 for more examples).
We call them the large-cool-core (LCC) class and the corona class. The gas of the
former class is primary of ICM origin, while the latter one is of ISM origin.
For 22 non-BCGs with $L_{\rm 1.4 GHz} > 10^{24}$ W Hz$^{-1}$ AGN, there are
only six confirmed corona detections, but most upper limits are high.

2) We emphasize that the above result is different from the well-known result
that almost every group or cluster with a strong large cool core  has a radio
AGN (e.g., Burns 1990; Eilek 2004). Our result also shows that the traditional
cool core / noncool core dichotomy is too simple. A better alternative is the
cool core distribution function with enclosed X-ray luminosity or gas mass
(e.g., the histogram in Fig. 1).

3) Small coronae, easily overlooked or misidentified as X-ray AGN at $z>$0.1, are
mini-cool-cores in groups and
clusters (Section 4, 5 and 6). They can trigger strong radio outbursts long before large
cool cores are formed. The triggered outbursts may destroy embryonic large
cool cores and thus provide another mechanism besides mergers to prevent formation of
large cool cores (see Burns et al. 2008). The outbursts triggered by coronae can
also inject extra entropy into the ICM and modify the ICM properties in systems
without large cool cores. For BCGs with strong radio AGN in our
sample, the corona fraction is at least comparable to that of large cool cores.

4) There are no groups with a luminous X-ray cool core ($L_{\rm 0.5-2 keV} > 10^{41.8}$ ergs s$^{-1}$)
hosting a strong radio AGN with $L_{\rm 1.4GHz} > 10^{24}$ W Hz$^{-1}$ (Section 4 and Fig. 1).
This is not observed in clusters ($kT >$ 2 keV). The absence of low-mass
systems with strong radio AGN creates a gap between the two classes at high
radio luminosities (Fig. 1). Although this result needs to be examined with a
larger, representative group sample (e.g., purely optically selected), it may
point to a greater impact of feedback on low-mass systems than clusters. We
suggest that some groups with a luminous cool core may have been transferred
into the corona class in a strong radio outburst.

5) In the LCC class, there is a general trend (albeit with large scatter)
that more luminous cool cores host more luminous radio AGN. We suspect that
the environmental boosting may play a role to create the trend. BCGs in weak cool
cores and noncool cores only have weak radio AGN (Fig. 1, 11 and Section 4).

6) Only $\sim$ 16\% of radio AGN ($L_{\rm 1.4GHz} > 10^{24}$ W Hz$^{-1}$) have
luminous X-ray AGN ($L_{\rm 0.5 - 10 keV} > 10^{42}$ ergs s$^{-1}$), while the
X-ray AGN fraction is even smaller for BCGs ($\sim$ 4\%). On the other hand,
$\geq$ 78\% of strong radio AGN have a confirmed cool core ($\geq$ 90\% for BCGs),
which implies their tight connection. For BCGs, all detected X-ray AGN
($L_{\rm 0.5 - 10 keV} > 10^{42}$ ergs s$^{-1}$) are also a radio AGN
($L_{\rm 1.4GHz} > 10^{24}$ W Hz$^{-1}$, Section 4.3).
Thus, a strong X-ray AGN of a BCG only emerges at the stage of its strong radio
activity, occasionally.

7) Luminous coronae may be able to power their radio AGN through Bondi accretion
(Section 7.1 and Fig. 10, also see Hardcastle et al. 2007), while the hot accretion
may not work for faint coronae in less massive galaxies. However, a complete inventory of cold gas in embedded
coronae is required to address the question of the accretion mode. We notice
that cold ISM and dust indeed exist in some coronae with strong radio AGN.

8) While coronae may trigger radio AGN, strong outbursts have to deposit
little energy inside coronae to keep them intact (also emphasized in S07). 
Thus, it is unclear whether coronae are decoupled from the radio feedback
cycle (Section 7.2). On the other hand, weak outbursts can provide gentle
heating required to offset cooling in coronae.
The existence of coronae around strong radio AGN in groups and clusters
also affects the properties of radio jets, e.g., extra pressure to decelerate jets
and maintain the collimation of jets. Its small size also allows the bulk of the
jet energy to transfer to the outskirts of the system.

9) We also present detailed analyses on coronae associated with \ga\ (in A3627)
and NGC 4709 (in the Centaurus cluster), as an example of luminous coronae
associated with a strong radio AGN (Section 5) and an example of faint coronae
(Section 6) respectively.

We want to further stress the importance of small X-ray cool cores like coronae.
There has been a lot of attention to understand large cool cores like Perseus's.
However, most of local massive galaxies (e.g., more luminous than $L_{*}$) are
not in large cool cores. Small X-ray cool cores like coronae may be more typical
gaseous atmosphere that actually matters in the evolution of massive galaxies to
explain their colors and luminosity function (see the beginning of Section 1).
There has also been a lot of discussions on ``bimodality'' of the cluster/group gas cores
(e.g., Cavagnolo et al. 2009), basically cool cores (with low entropy and power-law
distribution) and noncool cores (with high entropy and flat distribution). 
However, {\em one should not misunderstand the bimodality of the ICM entropy at $\geq$ 5 - 10 kpc
scales with the bimodality of the gas entropy at $\sim$ Bondi radius of the central
SMBH.}
The formal ``bimodality'' has been confirmed (e.g., Cavagnolo et al. 2009), while
we now know that clusters are identified as ``noncool core'' clusters at $\geq$ 5 - 10 kpc
scales can still have low-entropy gas around their BCGs. 
The properties of the gaseous atmosphere around the central SMBH are not necessarily
correlated with the gas properties at larger scales (e.g., $r \geq$ 5 - 10 kpc).

\acknowledgments

We are very grateful to Paul Nulsen, Megan Donahue, Christine Jones and Mark Voit
for comments and suggestions on an early draft of this paper.
We also want to thank Alexey Vikhlinin and Bill Forman for inspiring
discussions since the discovery of embedded coronae.
We also thank Craig Sarazin, Ken Cavagnolo and Greg Sivakoff for discussions.
We thank Ken Cavagnolo for providing his results on some cool-core clusters.
We thank Alexey Vikhlinin to provide the proprietary \chandra\ data of A3135,
A3532 and 3C~88. We also thank the referee for helpful comments.
This research has made use of the NASA/IPAC Extragalactic Database (NED)
which is operated by the Jet Propulsion Laboratory, California Institute of
Technology, under contract with the National Aeronautics and Space Administration.
The financial support for this work was provided by the NASA grant GO7-8081A, GO8-9083X and
the NASA LTSA grant NNG-05GD82G.

\vspace{-1cm}
\begin{table}
\begin{center}
\caption{The sample of 152 groups and clusters}
{\scriptsize
\begin{tabular}{ccc|ccc|ccc|ccc|ccc} \hline \hline
System & $z$\tablenotemark{a} & N\tablenotemark{b} & System & $z$\tablenotemark{a} & N\tablenotemark{b} & System & $z$\tablenotemark{a} & N\tablenotemark{b} & System & $z$\tablenotemark{a} & N\tablenotemark{b} & System & $z$\tablenotemark{a} & N\tablenotemark{b} \\ \hline

Centaurus & 0.0114 & 8 & A1367    & 0.0220 & 3 & IC~1262  & 0.0326 & 4 & A1736    & 0.0458 & 1 & A3266 & 0.0589 & 2 \\
NGC~1550 & 0.0124 & 4  & NGC~5171 & 0.0229 & 1 & A496     & 0.0329 & 3 & A1644    & 0.0473 & 2 & A3158 & 0.0597 & 3 \\
NGC~7619 & 0.0125 & 2  & A3581    & 0.0230 & 1 & A1314    & 0.0335 & 1 & NGC~326  & 0.0474 & 1 & A3128 & 0.0599 & 1 \\
IC~4296  & 0.0125 & 2  & NGC~5129 & 0.0230 & 2 & IC1880   & 0.0340 & 1 & A4059    & 0.0475 & 2 & A3125 & 0.0611 & 2 \\
A1060    & 0.0126 & 1  & Coma     & 0.0231 & 9 & UGC3957  & 0.0341 & 1 & A3556    & 0.0479 & 1 & AS405 & 0.0613 & 1 \\
NGC~6482 & 0.0131 & 1  & NGC~1132 & 0.0233 & 1 & NGC~6269 & 0.0348 & 1 & A3558    & 0.0480 & 1 & A3135 & 0.0623 & 1 \\
HCG42    & 0.0133 & 1  & A400     & 0.0244 & 1 & A1142    & 0.0349 & 1 & SC1329-313 & 0.0482 & 1 & A1795 & 0.0625 & 13 \\
Pavo     & 0.0137 & 1  & NGC~7386 & 0.0244 & 1 & A2063    & 0.0349 & 4 & A193     & 0.0486 & 1 & A2734 & 0.0625 & 1 \\
HCG62    & 0.0137 & 1  & UGC~2755 & 0.0245 & 1 & 2A0335+096 & 0.0349 & 3 & A3562    & 0.0490 & 1 & A1275 & 0.0637 & 1 \\
NGC~5419 & 0.0138 & 2  & 3C296    & 0.0247 & 1 & A2147    & 0.0350 & 1 & SC1327-312 & 0.0495 & 1 & A695  & 0.0687 & 2 \\
NGC~4782 & 0.0144 & 1  & NGC~6251 & 0.0247 & 2 & A2052    & 0.0355 & 2 & A2717    & 0.0498 & 2 & A3120 & 0.0690 & 1 \\
NGC~2563 & 0.0149 & 6  & NGC~4325 & 0.0257 & 1 & ESO~306-017 & 0.0358 & 2 & A3395S   & 0.0506 & 1 & A514  & 0.0713 & 1 \\
NGC~3402 & 0.0153 & 1  & HCG~51   & 0.0258 & 2 & A2151    & 0.0366 & 1 & A1377    & 0.0514 & 1 & A744  & 0.0729 & 1 \\
NGC~1600 & 0.0156 & 2  & 3C~442A  & 0.0263 & 4 & NGC~5098 & 0.0368 & 2 & A3391    & 0.0514 & 1 & A2462 & 0.0733 & 1 \\
A3627    & 0.0162 & 5  & MKW8     & 0.0270 & 1 & A576     & 0.0389 & 1 & A3528S   & 0.0530 & 1 & A3112 & 0.0752 & 5 \\
A262     & 0.0163 & 2  & UGC~5088 & 0.0274 & 1 & RBS540   & 0.0390 & 1 & Hydra A  & 0.0539 & 4 & A2670 & 0.0761 & 1 \\
NGC~507  & 0.0164 & 2  & NGC~6338 & 0.0274 & 1 & A3571    & 0.0391 & 1 & A754     & 0.0542 & 7 & A2029 & 0.0773 & 3 \\
NGC~315  & 0.0165 & 1  & NGC~4104 & 0.0282 & 1 & AS463    & 0.0394 & 2 & RXJ~1022+3830 & 0.0543 & 1 & RXJ~1159+5531 & 0.0808 & 1 \\
NGC~777  & 0.0167 & 1  & PKS2153-69 & 0.0283 & 1 & A1139    & 0.0398 & 1 & A85    & 0.0551 & 9 & A1650 & 0.0839 & 8 \\
3C~31    & 0.0170 & 1  & A539     & 0.0284 & 2 & A2657    & 0.0402 & 1 & A2626  & 0.0553 & 1 & A2597 & 0.0852 & 3 \\
3C~449   & 0.0171 & 1  & Ophiuchus & 0.0291 & 1 & A2572    & 0.0403 & 1 & A3532  & 0.0554 & 1 & A478  & 0.0881 & 11 \\
AWM7     & 0.0172 & 1  & RBS~461  & 0.0296 & 1 & A2107    & 0.0411 & 1 & A3667  & 0.0556 & 9 & A2142 & 0.0909 & 4 \\
NGC~7618 & 0.0173 & 3  & A4038    & 0.0300 & 2 & A2589    & 0.0414 & 4 & A2319  & 0.0557 & 1 & 3C388 & 0.0917 & 2 \\
UGC12491 & 0.0174 & 1  & A2199    & 0.0302 & 2 & NGC~2484 & 0.0428 & 1 & Cygnus A & 0.0561 & 11 & A2384 & 0.0943 & 1 \\
Perseus  & 0.0179 & 19 & 3C88     & 0.0302 & 1 & A119     & 0.0442 & 2 & AS1101 & 0.0564 & 1 & A2244 & 0.0968 & 1 \\
A194     & 0.0180 & 2  & ZW1615+35 & 0.0310 & 4 & A160     & 0.0447 & 1 & A133   & 0.0566 & 3 & PKS~0745-191 & 0.103 & 3 \\
NGC~533  & 0.0185 & 1  & ESO~552-020 & 0.0314 & 1 & A168     & 0.0450 & 2 & ESO~351-021 & 0.0571 & 1 & A1446 & 0.104 & 1 \\
NGC~741  & 0.0185 & 1  & A2634    & 0.0314 & 1 & MKW3S    & 0.0450 & 1 & A2256 & 0.0581 & 3 & RX J1852.1+5711  & 0.109 & 1 \\
MKW4     & 0.0200 & 1  & A1177    & 0.0316 & 1 & UGC~842  & 0.0452 & 1 & A3880 & 0.0581 & 1 & A562 & 0.110 & 1 \\
3C129.1  & 0.0210 & 2  & AWM4     & 0.0317 & 1 & A3376    & 0.0456 & 2 & A1991 & 0.0587 & 1 & A2220 & 0.110 & 1 \\
3C66B    & 0.0212 & 1  & A1185    & 0.0325 & 1 & & & & & & & & & \\

\hline \hline
\end{tabular}}
\vspace{-1cm}
\begin{flushleft}
\leftskip 35pt
\tablenotetext{a}{The redshift is from NASA/IPAC Extragalactic Database (NED)}
\tablenotetext{b}{The number of \chandra\ observations}
\end{flushleft}
\end{center}
\end{table}
\clearpage

\begin{table}
\begin{center}
\caption{Galaxies in the corona class with $L_{\rm 1.4 GHz} > 10^{24}$ W Hz$^{-1}$ AGN}
{\scriptsize
\begin{tabular}{cccccc} \hline \hline
Galaxy (Cluster) & $L_{\rm Ks}\tablenotemark{a}$ & $L_{\rm 1.4 GHz}\tablenotemark{b}$ & $L_{\rm 0.5-2 keV}\tablenotemark{c}$ & $kT$ (keV) & Note\tablenotemark{d} \\ \hline

IC~4296 (IC~4296) & 11.82 & 24.14 & 1.50$\pm$0.12 & 0.68$\pm$0.02 & BCG, $2.9\times10^{41}$ erg s$^{-1}$ AGN \\
3C~278 (NGC~4782) & 11.81 & 24.54 & 0.223$\pm$0.016 & 0.71$^{+0.04}_{-0.06}$ & BCG \\
ESO~137-006 (A3627) & 11.77 & 25.39 & 1.69$\pm$0.08 & 0.91$\pm$0.02 & BCG \\
ESO~137-007 (A3627) & 11.44 & 24.39 & $<0.13$ & & NAT, X-ray PS \\
NGC~315 (NGC~315) & 11.88 & 24.12 & 1.97$\pm$0.06 & 0.59$\pm$0.01 & BCG, $8.0\times10^{41}$ erg s$^{-1}$ AGN \\
3C~31 (3C~31) & 11.70 & 24.49 & 0.677$\pm$0.028 & 0.72$\pm$0.02 & BCG, $1.1\times10^{41}$ erg s$^{-1}$ AGN \\
3C~449 (3C~449) & 11.16 & 24.38 & 0.246$\pm$0.033 & 0.63$\pm$0.07 & BCG \\
NGC~1265 (Perseus) & 11.60 & 24.79 & 0.322$\pm$0.039 & 0.63$\pm$0.03 & NAT \\
NGC~547 (A194) & 11.74 & 24.45 & 0.824$\pm$0.021 & 0.66$\pm$0.02 & BCG, $1.7\times10^{41}$ erg s$^{-1}$ AGN \\
3C~129.1 (3C129.1) & 11.66 & 24.28 & 0.91$\pm$0.26 & 0.93$^{+0.32}_{-0.60}$ & BCG \\
3C~129 (3C129.1) & 11.50 & 24.75 & $<0.22$ & & NAT, X-ray PS \\
3C~66B (3C~66B) & 11.58 & 24.91 & 0.691$\pm$0.043 & 0.59$\pm$0.03 & BCG, $3.2\times10^{41}$ erg s$^{-1}$ AGN \\
NGC~3862 (A1367) & 11.52 & 24.77 & 0.14$^{+0.09}_{-0.05}$ & 0.65$^{+0.29}_{-0.09}$ & BCG \\
3C~75 (A400) & 12.06 & 24.90 & 0.986$\pm$0.079 & 0.70$^{+0.06}_{-0.09}$, 0.81$\pm$0.08 & BCG, two coronae added, $\sim 3\times10^{41}$ erg s$^{-1}$ AGN \\
NGC~7385 (NGC~7386) & 11.72 & 24.51 & 0.70$\pm$0.11 & 0.61$^{+0.06}_{-0.05}$ & NAT, $3.4\times10^{41}$ erg s$^{-1}$ AGN \\
UGC~2755 (UGC~2755) & 11.51 & 24.28 & 1.01$\pm$0.12 & 0.60$\pm$0.05 & BCG, $1.8\times10^{41}$ erg s$^{-1}$ AGN \\
3C~296 (3C~296) & 11.91 & 24.78 & 1.3$\pm$0.1 & 0.73$\pm$0.02 & BCG, $3.2\times10^{41}$ erg s$^{-1}$ AGN \\
NGC~6251 (NGC~6251) & 11.81 & 24.55 & 1.25$\pm$0.12 & 0.69$\pm$0.03 & BCG, $8.5\times10^{42}$ erg s$^{-1}$ AGN \\
3C~442A (3C~442A) & 11.78 & 24.73 & 3.21$\pm$0.20 & 0.96$\pm$0.03 & BCG (two galaxies added), $r\sim$ 25 kpc core \\
PKS~2152-69 (PKS~2152-69) & 11.52 & 25.72 & 0.77$\pm$0.20 & 0.75$^{+0.08}_{-0.11}$ & BCG, $9.0\times10^{42}$ erg s$^{-1}$ AGN \\
3C~88 (3C~88) & 11.40 & 25.03 & 2.75$\pm$0.30 & 0.77$\pm$0.05 & BCG, $r\sim$ 25 kpc core, $\sim 8\times10^{41}$ erg s$^{-1}$ AGN \\
NGC~6109 (ZW~1615+35) & 11.49 & 24.64 & 0.754$\pm$0.102 & 0.58$\pm$0.05 & NAT \\
3C~465 (A2634) & 11.91 & 25.21 & 2.18$\pm$0.35 & 1.01$\pm$0.03 & BCG \\
NGC~6051 (AWM4) & 11.85 & 24.17 & 0.21$\pm$0.03 & 1.32$^{+0.16}_{-0.12}$ & BCG \\
GIN~190 (A496) & 11.27 & 24.26 & $<0.12$ & & NAT, X-ray PS \\
2MASX~J03381409+1005038 (2A0335+096) & 11.71 & 24.06 & $<0.23$ & & NAT, $7.6\times10^{41}$ erg s$^{-1}$ AGN \\
PKS~0427-53 (AS463) & 11.92 & 25.32 & 0.791$\pm$0.070 & 0.89$^{+0.10}_{-0.15}$ & BCG \\
NGC~2484 (NGC~2484) & 11.96 & 25.05 & 1.5$\pm$0.4 & 0.70$^{+0.12}_{-0.20}$ & BCG, $1.7\times10^{42}$ erg s$^{-1}$ AGN \\
UGC~583 (A119) & 11.76 & 24.79 & 0.73$\pm$0.09 & 0.68$\pm$0.06 & NAT \\
CGCG~384-032 (A119) & 11.50 & 24.72 & $<0.21$ & & NAT, weak PS \\ 
GIN~049 (A160) & 11.71 & 24.68 & 0.47$\pm$0.07 & 1.17$\pm$0.13 & BCG \\ 
PMN~J1326-2707 (A1736) & 11.52 & 24.08 & $<$0.34 & & weak PS \\
PGC~018297 (A3376) & 11.52 & 24.28 & 0.32$\pm$0.13 & 0.60$^{+0.16}_{-0.15}$ & BCG, code 3 \\
NGC~326 (NGC~326) & 11.96 & 24.98 & 0.740$\pm$0.056 & 0.66$\pm$0.04 & BCG, $\sim 1.5\times10^{41}$ erg s$^{-1}$ AGN \\
PKS~0625-545 (A3395S) & 11.76 & 25.31 & 1.5$\pm$0.3 & 1.59$^{+0.20}_{-0.23}$ & BCG \\
PKS~0625-53 (A3391) & 12.07 & 25.60 & 1.05$\pm$0.24 & 0.90$\pm$0.13 & BCG \\
PGC~025672 (A754) & 11.57 & 24.63 & $<3.6$ & & NAT, 1.4$\times10^{43}$ erg s$^{-1}$ AGN \\
PGC~025790 (A754) & 11.63 & 24.41 & $<0.90$ & & NAT, 3.3$\times10^{41}$ erg s$^{-1}$ AGN \\
2MASX~J09101737-0937068 (A754) & 11.74 & 24.49 & 0.97$\pm$0.30 & 0.95$^{+0.22}_{-0.19}$ & NAT, code 3 \\
PKS~1254-30 (A3532) & 12.02 & 24.91 & 0.84$\pm$0.30 & 1.00$^{+0.36}_{-0.42}$ & BCG, code 3 \\
PGC~064228 (A3667) & 11.32 & 24.60 & $<0.30$ & & NAT, $5.2\times10^{42}$ erg s$^{-1}$ AGN \\
PKS~0326-536 (A3125) & 11.43 & 24.50 & $<0.39$ & & NAT, weak PS \\
2MASX~J03275206-5326099 (A3125) & 11.52 & 24.29 & $<0.60$ & & NAT, weak PS \\
2MASX~J22275066-3033431 (A3880) & 11.25 & 24.25 & $<2.0$ & & NAT, $1.1\times10^{43}$ erg s$^{-1}$ AGN \\
PKS~0429-61 (A3266) & 10.70 & 25.14 & $<0.14$ & & NAT, no X-ray \\
PKS~0332-39 (A3135) & 11.60 & 25.15 & 1.37$\pm$0.35 & 0.86$^{+0.17}_{-0.08}$ & BCG, $\sim2\times10^{41}$ erg s$^{-1}$ AGN \\
4C +32.26 (A695) & 11.76 & 24.92 & 0.66$\pm$0.12 & 0.95$^{+0.10}_{-0.15}$ & BCG, $r\sim$ 20 kpc core, $3.4\times10^{41}$ erg s$^{-1}$ AGN \\
2MASX~J04483058-2030478 (A514) & 11.63 & 24.93 & 0.73$\pm$0.19 & 0.93$^{+0.18}_{-0.14}$ & NAT, code 3 \\
2MASX~J04481046-2024574 (A514) & 11.48 & 24.24 & $<1.5$ & & NAT, $1.1\times10^{42}$ erg s$^{-1}$ AGN \\
2MASX~J04480299-2026384 (A514) & 11.32 & 24.16 & $<0.47$ & & NAT, $1.2\times10^{41}$ erg s$^{-1}$ AGN \\
PKS~2236-17 (A2462) & 11.85 & 25.38 & 1.43$\pm$0.11 & 0.82$\pm$0.05 & BCG, $\sim2\times10^{41}$ erg s$^{-1}$ AGN \\
B2~1556+27 (A2142) & 11.36 & 24.37 & $<0.60$ & & NAT, $2.1\times10^{42}$ erg s$^{-1}$ AGN \\
4C +58.23 (A1446) & 11.94 & 25.36 & 1.09$\pm$0.22 & 1.00$^{+0.27}_{-0.16}$ & BCG \\
4C +69.08 (A562) & 11.94 & 25.62 & 1.35$\pm$0.27 & 0.75$\pm$0.10 & BCG \\
SBS 1638+538 (A2220) & 12.26 & 25.31 & 1.64$\pm$0.33 & 0.96$^{+0.12}_{-0.28}$ & BCG, code 3 \\

\hline \hline 
\end{tabular}}
\vspace{-1cm}
\begin{flushleft} 
\leftskip 35pt
\tablenotetext{a}{2MASS $K_{\rm s}$ band luminosity of the galaxy as shown as log(L$_{\rm Ks}$/L$_{\odot}$), $M_{K\odot}$ = 3.39 mag.}
\tablenotetext{b}{1.4 GHz luminosity of the galaxy as shown as log(L$_{\rm 1.4 GHz}$/W Hz$^{-1}$) from the NRAO VLA Sky Survey (NVSS) or the Sydney University Molonglo Sky Survey (SUMSS), assuming a spectral index of -0.8.}
\tablenotetext{c}{The rest-frame 0.5 - 2 keV luminosity in unit of 10$^{41}$ ergs s$^{-1}$}
\tablenotetext{d}{BCGs are marked. Most of non-BCGs are NAT sources. The AGN luminosity (measured in the rest-frame 0.5 - 10 keV) is listed if it is higher than 10$^{41}$ ergs s$^{-1}$. ``code 3'' refers to ``soft X-ray sources'' defined in S07 and Section 3.}
\end{flushleft} 
\end{center}
\end{table}

\begin{table}
\begin{center}
{\small
\caption{Fractions of X-ray AGN and cool cores}
\begin{tabular}{lcc} \hline \hline
Sample & Fraction of $L_{\rm 0.5-10 keV} > 10^{42}$ ergs s$^{-1}$ AGN & Fraction of X-ray cool cores$^{a}$ \\ \hline

BCGs with $L_{\rm 1.4 GHz} > 10^{24}$ W Hz$^{-1}$ & 7/52 & 52/52 \\
BCGs with $L_{\rm 1.4 GHz} > 10^{23}$ W Hz$^{-1}$ & 7/81 & $>$ 79/81 \\
All $L_{\rm 1.4 GHz} > 10^{24}$ W Hz$^{-1}$ galaxies & 12/74 & $>$ 58/74 \\
All BCGs & 7/161 & $>$ 145/161 \\

\hline \hline
\end{tabular}
\vspace{-1cm}
\tablenotetext{a}{The fractions of coronae are $\sim$ 60\% in the first sample and
$\geq$ 50\% in the other three (Section 4).}
}
\end{center}
\end{table}

\begin{table}
\begin{center}
{\small
\caption{The coronae of \ga\ and NGC~4709}
\begin{tabular}{lcc} \hline \hline
Property$^{a}$ & \ga\ & NGC~4709 \\ \hline

 Cluster ($z$) & A3627 (0.0162) & Centaurus (0.0114) \\
 log(L$_{\rm Ks}$/L$_{\odot}$) & 11.77 & 11.27 \\
 log(L$_{\rm 1.4 GHz}$/W Hz$^{-1}$) & 25.39 & $<$20.88 \\
 $r_{\rm cut}$ (kpc) & 4.24$\pm$0.20 & $\sim$ 2.0 \\
 $kT$ (keV) &     0.91$\pm$0.02$^{b}$ & 0.78$\pm$0.09 \\
  O (solar) & 0.52$^{+0.63}_{-0.40}$ & (0.8) \\
 Ne (solar) & 4.14$^{+2.38}_{-1.51}$ & (0.8) \\
 Mg (solar) & 2.27$^{+1.19}_{-0.70}$ & (0.8) \\
 Si (solar) & 1.66$^{+0.82}_{-0.48}$ & (0.8) \\
  S (solar) & 2.76$^{+1.55}_{-0.99}$ & (0.8) \\
 Fe (Ni) (solar) & 1.13$^{+0.49}_{-0.28}$ & (0.8) \\
 $f_{\rm 0.5-2 keV, thermal, obs}$ (ergs cm$^{-2}$ s$^{-1}$) & (1.77$\pm$0.09)$\times10^{-13}$ & (9.5$\pm$1.8)$\times10^{-15}$ \\
 $L_{\rm 0.5-2 keV, thermal}$ (ergs s$^{-1}$) & (1.69$\pm$0.08) $\times10^{41}$ & (1.89$\pm$0.36) $\times10^{39}$ \\
 $L_{\rm bol, thermal}$ (ergs s$^{-1}$) & (2.76$\pm$0.14) $\times10^{41}$ & (3.1$\pm$0.6) $\times10^{39}$ \\
 $L_{\rm 0.3-8 keV, LMXB}$ ($r<r_{\rm cut}$) (ergs s$^{-1}$) & (5.2$\pm1.1) \times10^{40}$ & $\sim 5.3\times10^{39}$ \\
 $L_{\rm 0.3-8 keV, LMXB}$ ($r=r_{\rm cut} - 11.1$ kpc) (ergs s$^{-1}$) & (4.1$\pm0.8) \times10^{40}$  & - \\
 $L_{\rm 0.3-10 keV, nucleus}$ (ergs s$^{-1}$) & $< 4\times10^{39}$ & $< 5\times10^{38}$ \\
 $n_{\rm e, center}$ (cm$^{-3}$) & $\sim$ 0.5 & $\sim$ 0.034 \\
 $t_{\rm cooling, center}$ (Myr) & $\sim$ 7 & $\sim$ 130 \\
 $M_{\rm gas}$ (M$_{\odot}$) & (1.78$\pm$0.15) $\times10^{8}$ & $\sim 2.9\times10^{6}$ \\

\hline \hline
\end{tabular}
\vspace{-1cm}
\tablenotetext{a}{The energy bands are all measured in the rest-frame.}
\tablenotetext{b}{See Fig. 7 for \ga's temperature profile.}
}
\end{center}
\end{table}

\begin{figure}
\vspace{-2cm}
\centerline{\includegraphics[height=0.9\linewidth,angle=270]{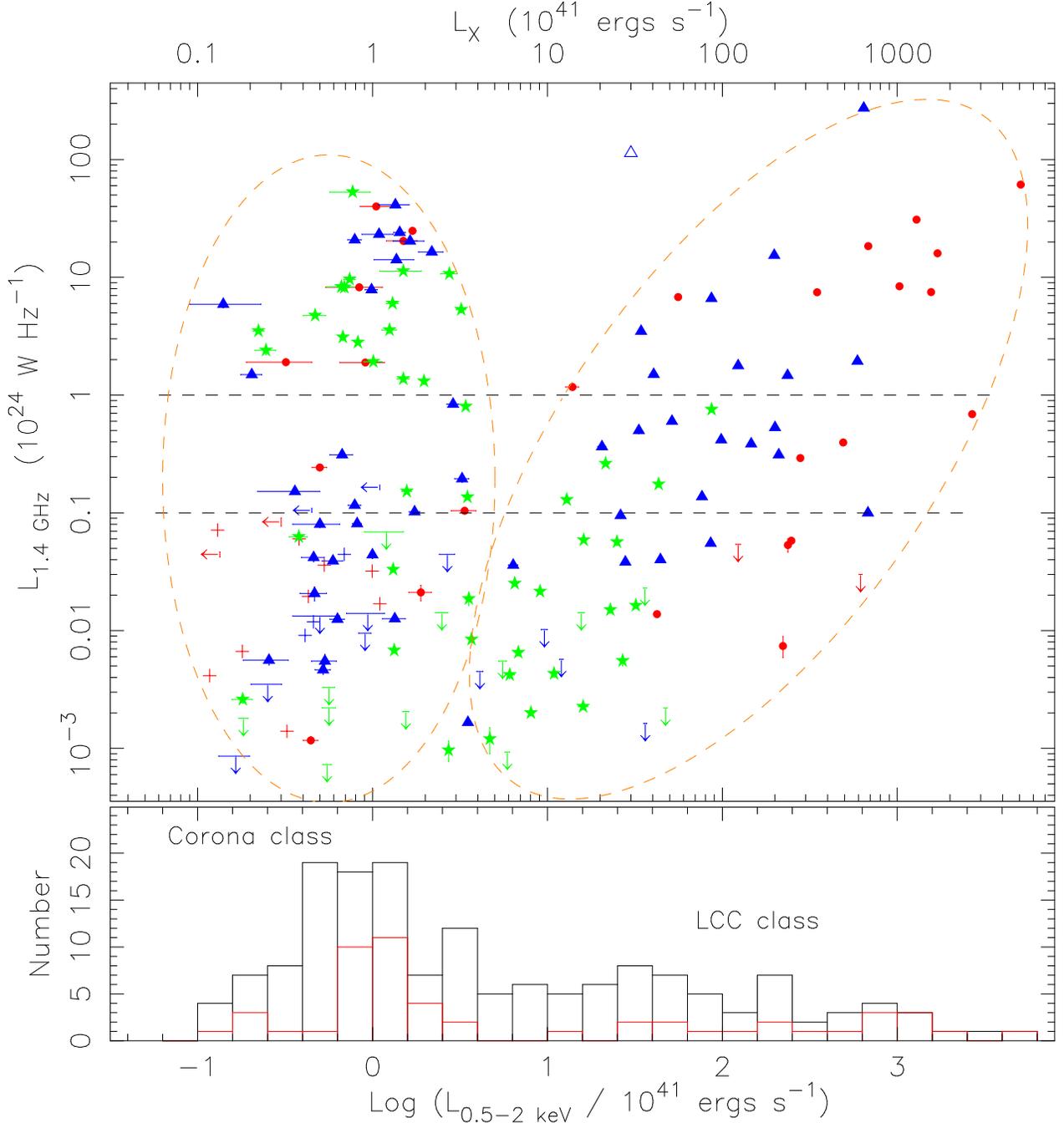}}
  \caption{The upper panel shows the rest-frame 0.5-2 keV luminosity of the cool core
(within a radius where the cooling time is 4 Gyr) of the BCG vs. the 1.4 GHz
luminosity of the BCG. Red filled circles are for $kT>$4 keV clusters. Blue triangles are
for $kT$=2-4 keV poor clusters. Green stars are for $kT<$2 keV groups.
Crosses represent upper limits in both axes.
The horizontal dashed lines are $L_{\rm 1.4GHz} = 10^{24}$ and $10^{23}$ W Hz$^{-1}$.
The lower panel shows the histogram for all BCGs (upper limits included)
with two classes marked, while the histogram in red is for $L_{\rm 1.4GHz} > 10^{24}$ W Hz$^{-1}$
BCGs. This histogram can be regarded as a raw {\bf cool-core distribution function}. 
At least three interesting results are revealed in this plot.
First, there are two classes of BCG cool cores (shown in
orange ellipses): the large-cool-core (LCC) class and the corona class.
Their dividing line is $\sim 4\times10^{41}$ ergs s$^{-1}$.
Above $L_{\rm 1.4GHz}$ of 10$^{23}$ W Hz$^{-1}$, every BCG has a confirmed cool core,
either in the LCC class or in the corona class, except for two BCGs with
high upper limits. In this work, most $L_{\rm 1.4GHz} > 10^{24}$ W Hz$^{-1}$
BCGs (33 out of 52) are in the corona class.
Second, there is a general trend (with large scatter) in the LCC class.
More luminous cool cores generally host more luminous radio AGN, or the LCC
class is tilted. The slope from the BCES Orthogonal fit is 1.91$\pm$0.20.
Third, there are no groups with a luminous cool core ($\geq 6\times10^{41}$ ergs s$^{-1}$)
that host a radio AGN more luminous than 10$^{24}$ W Hz$^{-1}$.
This is not observed in clusters (blue and red points).
The absence of luminous cool cores with strong radio AGN in low-mass systems
also causes a gap between the two classes at high radio luminosities.
Cygnus A is excluded in this plot as its position (7.7$\times10^{43}$ ergs s$^{-1}$,
1.2$\times10^{28}$ W Hz$^{-1}$) is so far away from those of others.
The only other source that is far from the two ellipses is 3C~388 (the open blue triangle between
two classes), which is also a FR-II galaxy like Cygnus A (Kraft et al. 2006).
}
\end{figure}

\begin{figure}
\centerline{\includegraphics[height=1.2\linewidth]{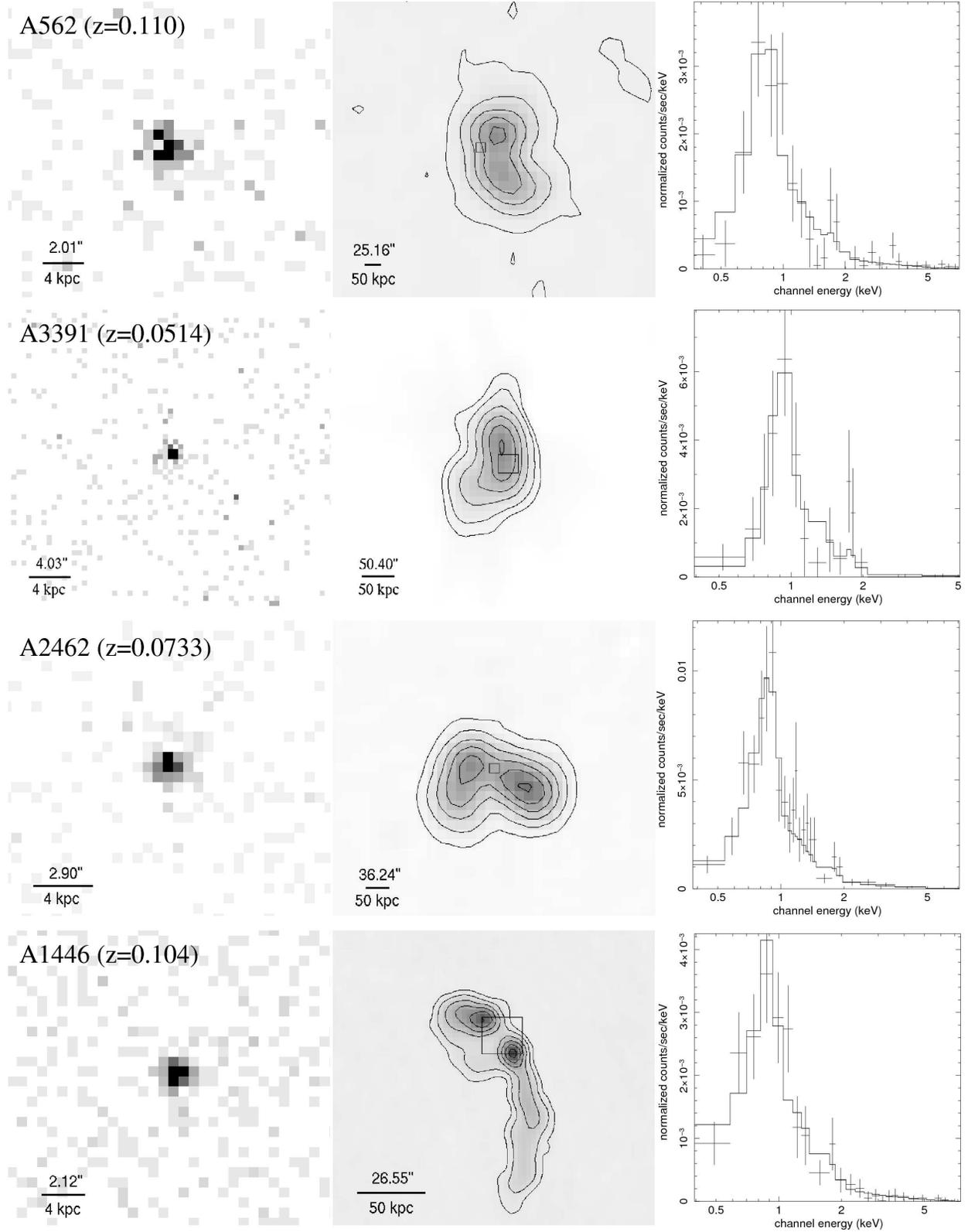}}
  \caption{Four of the five most luminous radio AGN in the corona class (the 3rd
most luminous one, \ga, is discussed in Section 5 and shown in Fig. 6): the BCGs of A562,
A3391, A2462 and A1446. All clusters were previously considered ``noncool core'' or weak cool core
clusters with central ICM cooling time of 9 Gyr, 17 Gyr, 9 Gyr and 6 Gyr respectively.
The left panel shows the 0.5 - 3 keV \chandra\ unbinned image.
Coronae are usually hardly resolved at their redshifts.
The middle panel shows the radio image and contours from
NVSS, FIRST or SUMSS. The small box shows the region of the left panel. These radio
sources have luminosities of 1.4 - 2.6 times Perseus's ($L_{\rm 1.4 GHz} = 1.6\times10^{25}$ W Hz$^{-1}$).
The right panel shows the spectrum of the corona. The iron L-shell hump is significant in all
cases, unambiguously confirming the existence of thermal gas.
}
\end{figure}
\clearpage

\begin{figure}
\centerline{\includegraphics[height=1.2\linewidth]{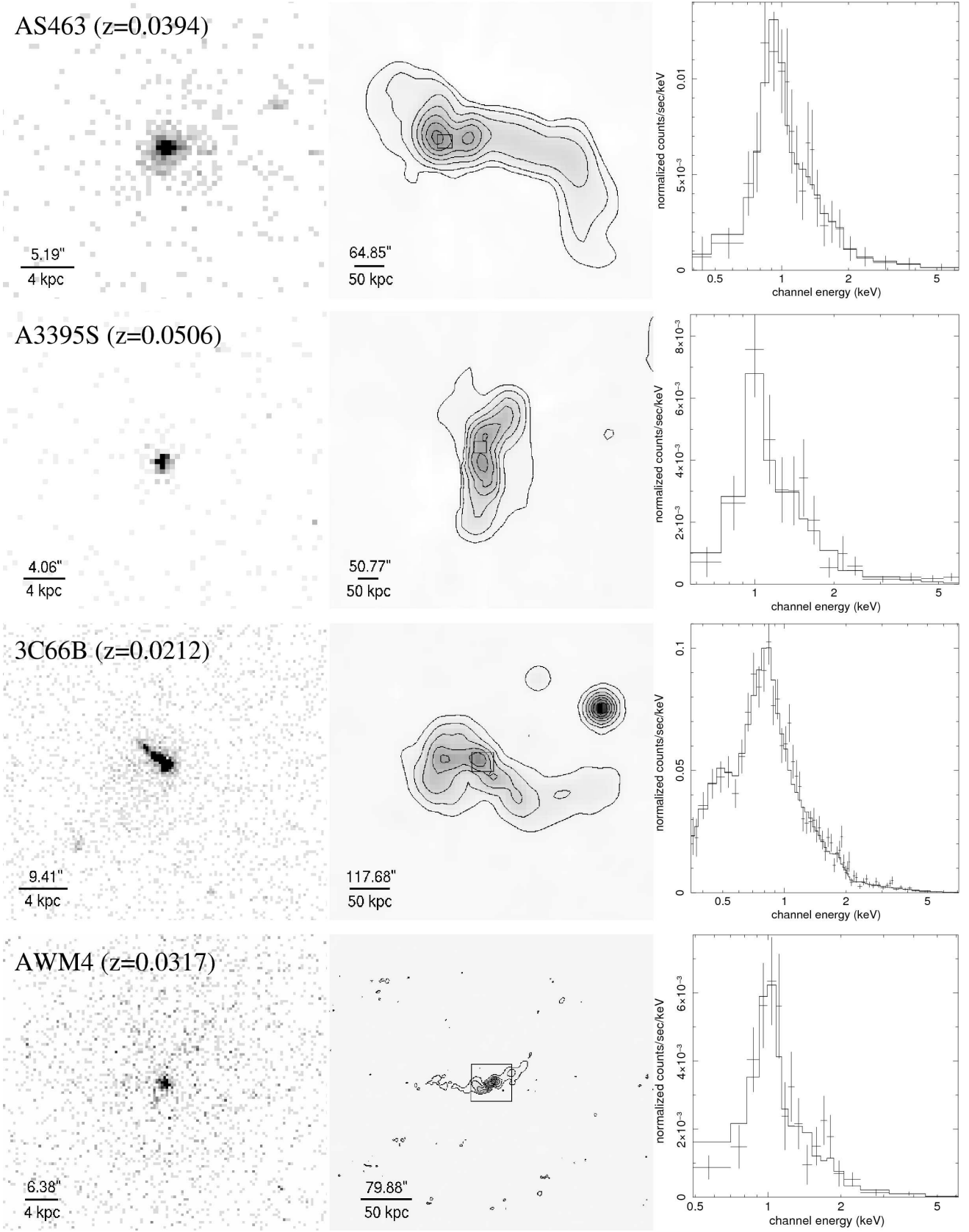}}
\vspace{0.2cm}
  \caption{Similar to Fig. 2, but for BCGs in AS463 and A3395S, the sixth and
seventh most luminous radio AGN (both are 1.3 times Perseus's) in the corona class.
Both were previously considered ``noncool core''  clusters with central ICM cooling
time of 10 Gyr and 15 Gyr respectively.
AS463's corona is clearly resolved owing to its relative
proximity and sufficient \chandra\ exposure (58 ks).
We also show an example of a small corona associated with a group's BCG,
3C~66B with a radio AGN that has half the luminosity of Perseus's. Despite
its bright nuclear source and X-ray jet, a corona is clearly resolved and the
iron L-shell hump is significant (also see Hardcastle et al. 2001; Croston et al. 2003).
The last example (AWM4) was once considered as a puzzle as it hosts a radio
AGN but lacks a large cool core from the \xmm\ data (Gastaldello et al. 2008).
The new \chandra\ data reveal that it is just another example of a small
corona associated with a radio AGN (that is much weaker than the other examples
in Fig. 2 and 3).
}
\end{figure}
\clearpage

\begin{figure}
\centerline{\includegraphics[height=0.5\linewidth,angle=270]{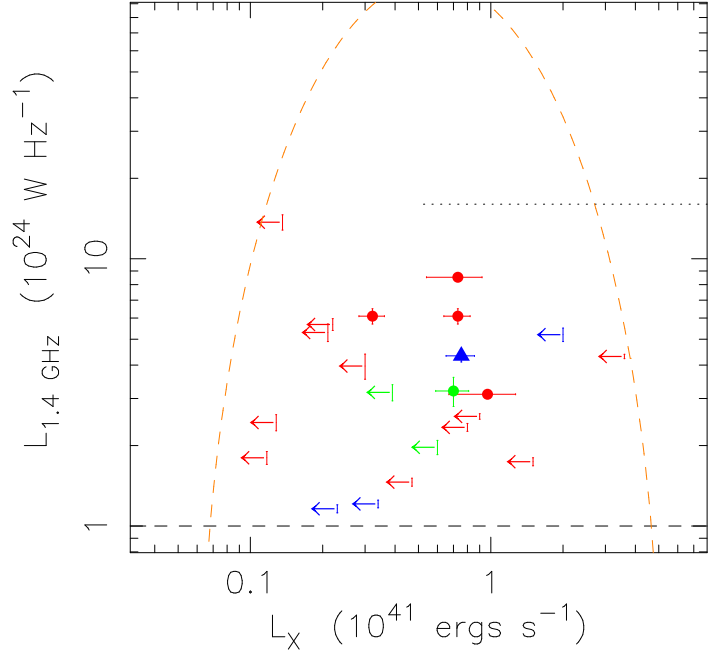}}
  \caption{The same plot as Fig. 1 but for non-BCGs with
$L_{\rm 1.4GHz} > 10^{24}$ W Hz$^{-1}$ (the dashed line). The same
orange ellipse for the corona class and the dotted line as Fig. 1 are shown.
Above the $L_{\rm 1.4GHz, cut}$, BCGs outnumbers non-BCGs, 52 vs. 22,
because of the limited FOV of \chandra\ (often centered on BCGs) and the
enhanced radio AGN activity of BCGs. Most non-BCGs do not have confirmed
coronae but most upper limits are high (see Fig. 5).
}
\end{figure}

\begin{figure}
\centerline{\includegraphics[height=1.0\linewidth,angle=270]{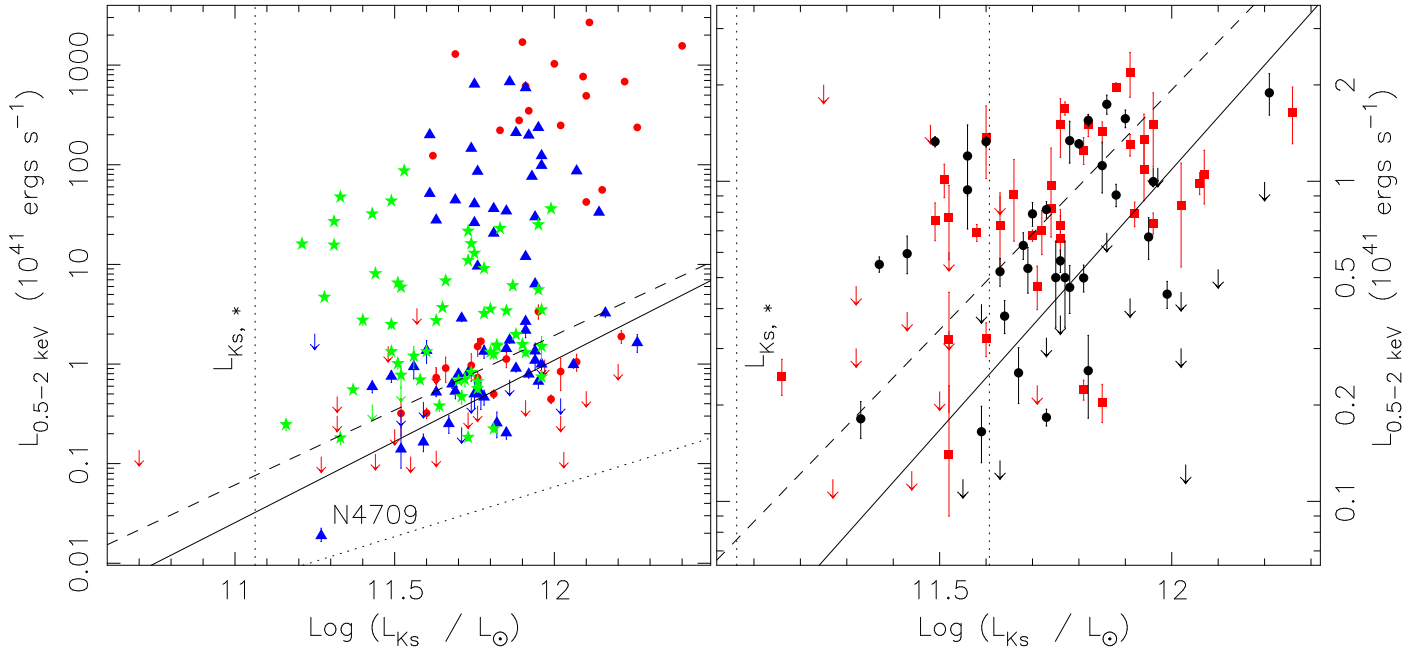}}
  \caption{{\bf Left}: 0.5-2 keV luminosity of the cool core (within a radius
where the cooling time is 4 Gyr) of the galaxy (including both BCGs and non-BCGs)
vs. the 2MASS $K_{\rm s}$ luminosity of the galaxy. Red points are for $kT>$4 keV clusters.
Blue triangles are for $kT$=2-4 keV poor clusters. Green stars are for $kT<$2 keV
groups. For coronae, the LMXB + nuclear emission has been subtracted as a power-law
component is always included in spectral fits.
The solid line is the best fit for both detections and upper limits of coronae
in rich clusters from S07, while the dashed line is the best fit for only detections
from S07. Large cool cores that are mostly composed of ICM form a different
population with higher X-ray luminosities than coronae that are composed of ISM.
$L_{\rm Ks, *}$ is marked by the vertical
dotted line. The position of NGC~4709 is marked, which is the faintest
corona known for galaxies more luminous than $L_{\rm Ks, *}$ (see Section 6).
The dotted line is the expected emission from cataclysmic variables and
coronally active stars (Revnivtsev et al. 2008).
As discussed in Section 6, the real contribution is even much smaller as
we always used the immediate local background that also includes much
stellar emission.
{\bf Right}: The same plot as the left one, but only for sources with
$L_{\rm 0.5-2 keV} < 2.5\times10^{41}$ ergs s$^{-1}$ (excluding
small cool cores). The red squares and upper limits are for
$L_{\rm 1.4 GHz} > 10^{24}$ W Hz$^{-1}$ galaxies, while the black points
and upper limits are for $L_{\rm 1.4 GHz} < 10^{24}$ W Hz$^{-1}$ galaxies.
At $L_{\rm Ks} > 3.5 L_{\rm Ks,*}$ (the dotted line), radio luminous
AGN are generally more luminous in X-rays than radio faint AGN
(medians: $\sim 9\times10^{40}$ ergs s$^{-1}$ vs. $\sim 5.5\times10^{40}$ ergs s$^{-1}$).
}
\end{figure}
\clearpage

\begin{figure}
\centerline{\includegraphics[height=0.6\linewidth]{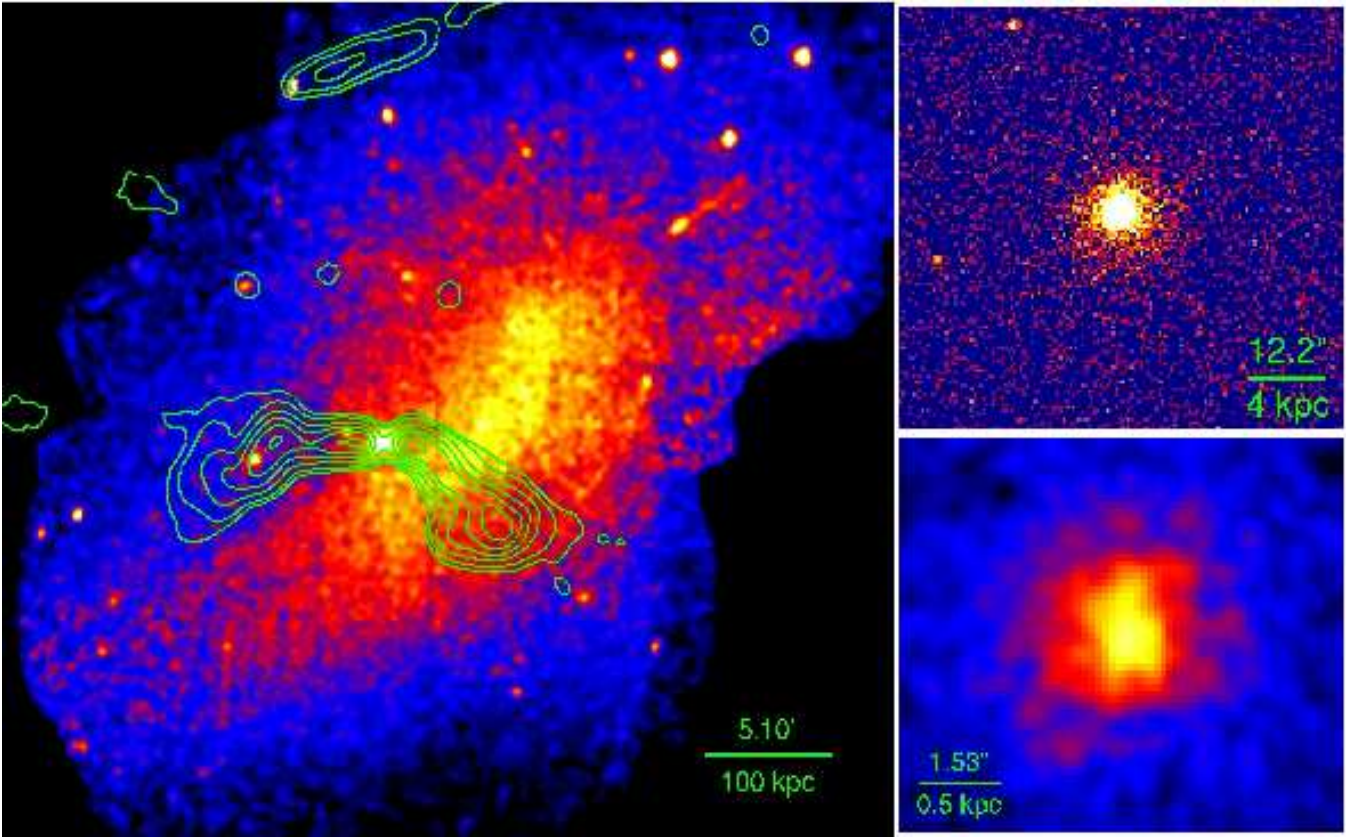}}
  \caption{843 MHz contours of \ga\ (from SUMSS, in green) overlaid on the 0.5 - 2 keV
\xmm\ mosaic image of A3627. Evidence indicative of interactions between radio plasma and the
ICM is present, including the positional coincidence of the southern boundary of the eastern
lobe with the eastern sharp edge of the ICM core and the expanding of the western
radio lobe beyond the bright ICM core.
The small right panel at the top shows the 0.5 - 5 keV \chandra\ image centered on \ga.
The nearby point sources show the size of the local PSF. \ga's corona is more
extended to the south, which implies its motion to the north and is consistent
with the bending of the radio lobes. The small panel at the bottom shows the
central region of the corona, after applying the ACIS Subpixel Event Repositioning tool
(Li et al. 2004). The core is elongated in the north-south, which is perpendicular
to the jet directions.
}
\end{figure}
\clearpage

\begin{figure}
\centerline{\includegraphics[height=0.72\linewidth]{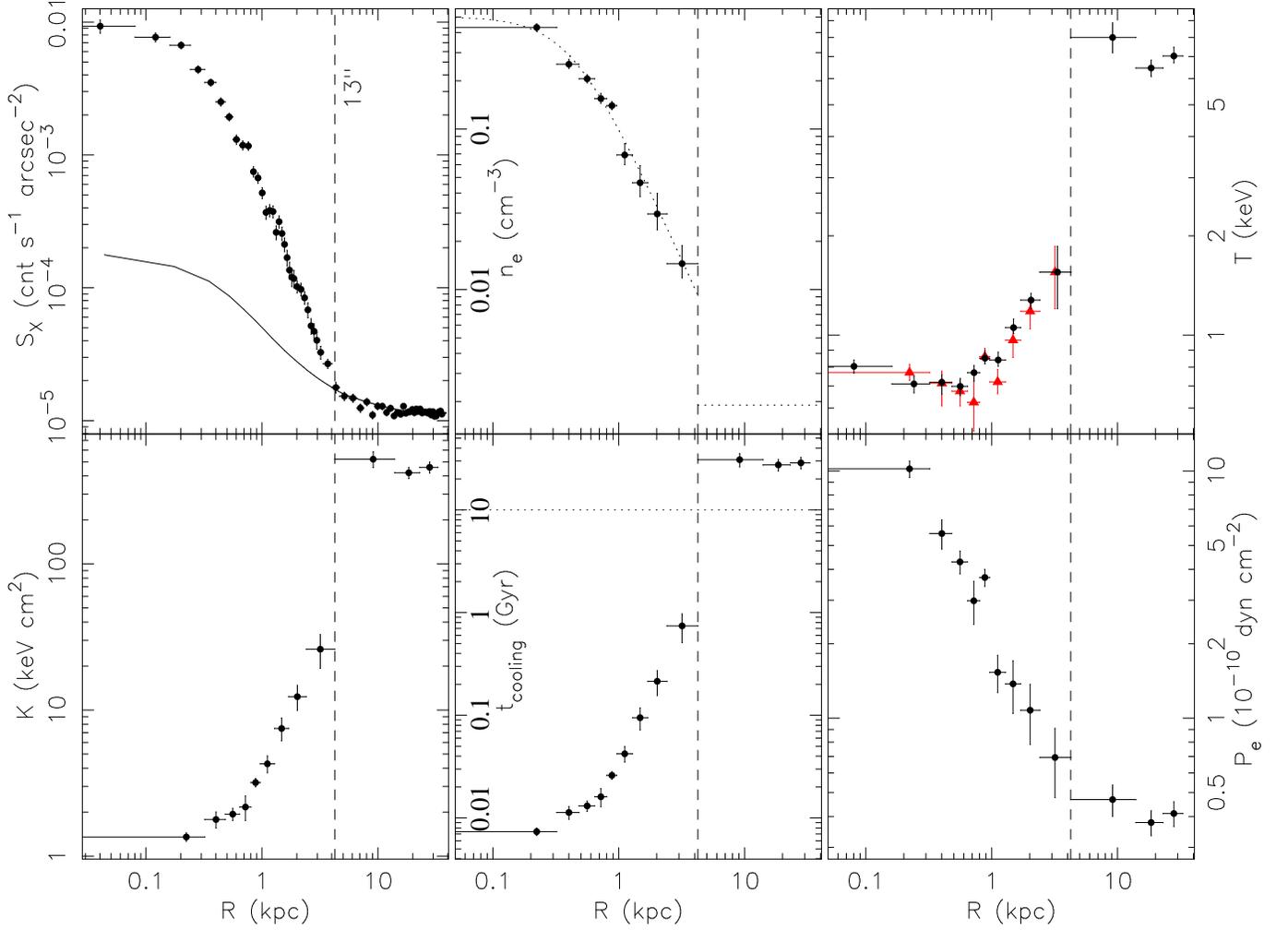}}
  \caption{Properties of \ga's corona and its surroundings are shown.
The {\em upper panel} shows the 0.5 - 2 keV surface brightness profile, the
electron density profile and the temperature profile. The solid line in the
left plot is the predicted LMXB light + the local background, which well describes the
surface brightness profile beyond $r_{\rm cut}$ (the vertical dashed line, 13$''$).
The dotted line in the middle plot is the best-fit $\beta$-model of the density
profile, while the horizontal dotted line is the average surrounding ICM density.
The red triangles in the right represent the deprojected temperature values.
The {\em lower panel} shows the entropy, cooling time and pressure profiles.
The whole corona region has low entropy and short cooling time ($<$ 1 Gyr).
Extrapolating the coronal gas pressure to the edge, the electron pressure ratio
across the edge is 1.16$\pm$0.51, which is consistent with pressure equilibrium.
}
\end{figure}

\begin{figure}
\centerline{\includegraphics[height=0.5\linewidth,angle=270]{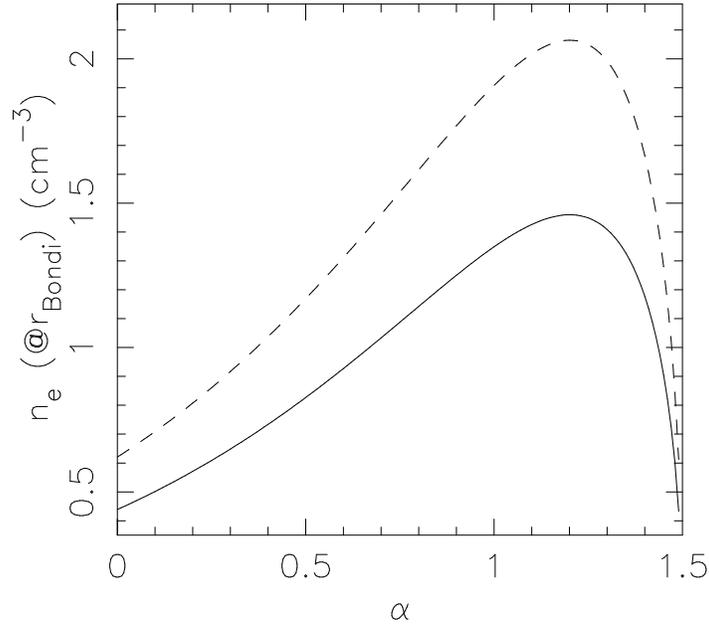}}
  \caption{The electron density of \ga's gas at the Bondi radius
(62 pc) vs. the slope of the density profile ($n_{\rm e} \propto r^{-\alpha}$).
The solid line is the relation for the observed normalization of the innermost
bin. No PSF correction was done. When $\alpha$=0, the density goes back to
the average value as shown in Fig. 7.
The dashed line is for the relation when the normalization
is doubled, which should over-estimate the PSF correction.
}
\end{figure}

\begin{figure}
\centerline{\includegraphics[height=0.45\linewidth,angle=270]{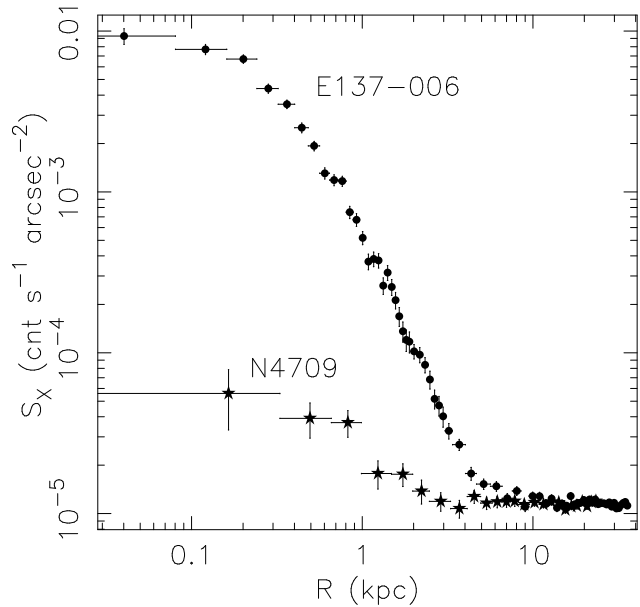}}
  \caption{0.5 - 2 keV surface brightness profiles of a faint embedded corona
(NGC~4709) vs. a luminous embedded corona (\ga). Their local background
has been matched to the same level. For NGC~4709's
profile, the LMXB light is subtracted (with the assumption to follow the
optical light) as it contributes half of the 0.5 - 2 keV emission.
These two examples show that the embedded coronae have wide ranges of the X-ray
luminosities (2$\times10^{39}$ ergs s$^{-1}$ to 2$\times10^{41}$ ergs s$^{-1}$),
central densities (0.03 cm$^{-3}$ to 0.5 cm$^{-3}$) and gas masses
(0.02 - 2$\times10^{8}$ M$_{\odot}$) (Table 4).
}
\end{figure}
\clearpage

\begin{figure}
\centerline{\includegraphics[height=0.9\linewidth,angle=270]{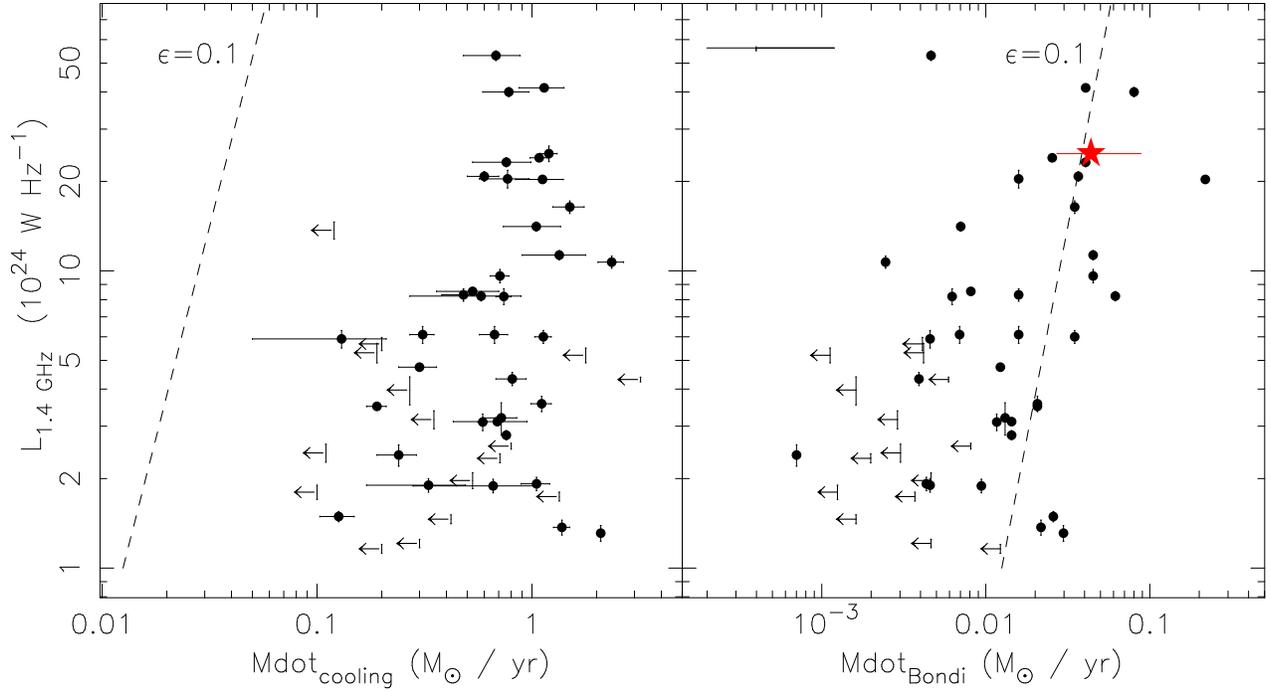}}
  \caption{{\bf Left}: the mass deposition rate from cooling (see $\S$7.1)
vs. the 1.4 GHz luminosity for coronae associated with strong radio AGN (i.e., the upper
portion of the corona class, Fig. 1). The dashed line shows the required SMBH
mass accretion rate to drive radio outflows for a mass-energy conversion efficiency
of 10\%. The relation between the radio luminosity and the mechanical power
of the radio outflows is from B\^{i}rzan et al. (2008), which has a large
scatter and also likely under-estimates the mechanical power.
This plot shows that in principle, only a small
fraction of the cooled coronal gas is needed to provide enough fuel to power radio AGN.
{\bf Right}: the estimated Bondi accretion rate vs. the 1.4 GHz luminosity
for coronae associated with strong radio AGN (see Section 7.1 for detail).
The assumed central entropy is 1.5 keV cm$^{2}$. The large errorbar shows the typical
range of $\dot{M}_{\rm Bondi}$ for the assumed entropy of 0.5 - 3 keV cm$^{2}$.
The dashed line is the same as the one in the left plot. The red star is \ga. As discussed
in Section 7.1, some luminous coronae of massive galaxies may be able to power their
radio AGN through Bondi accretion.
}
\end{figure}

\begin{figure}
\centerline{\includegraphics[height=0.92\linewidth,angle=270]{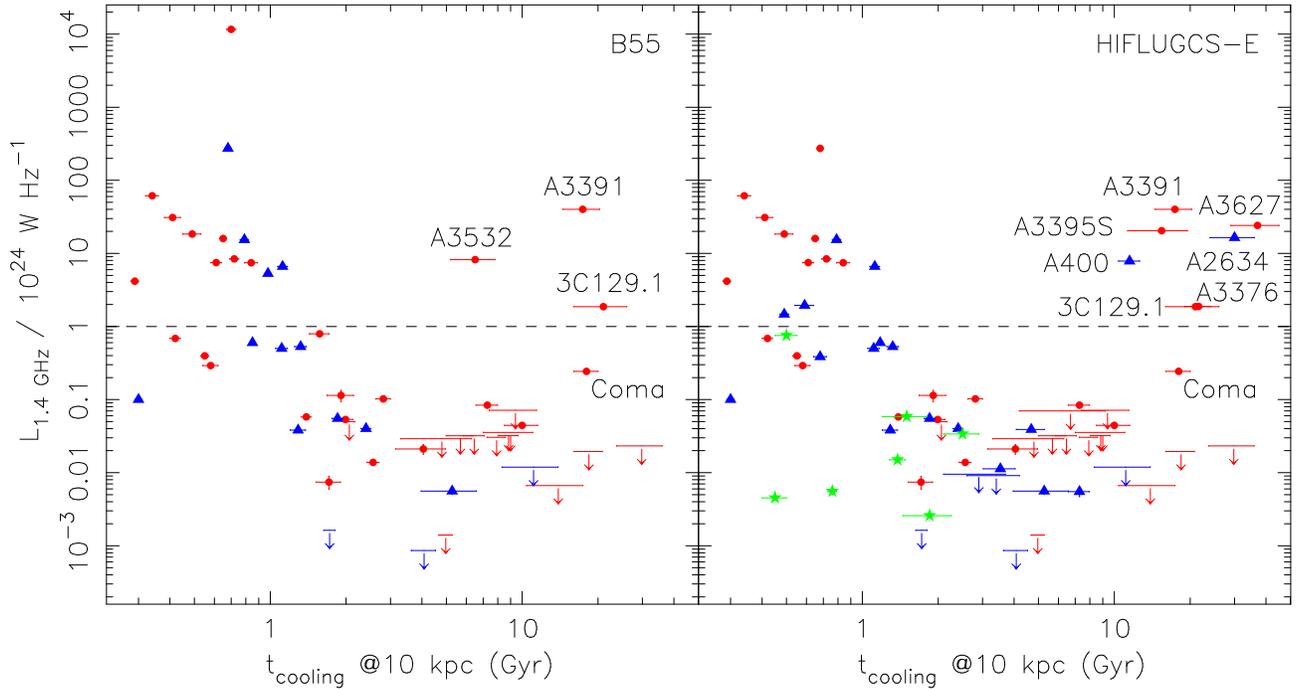}}
  \caption{Cooling time at 10 kpc radius of the BCG vs. the 1.4 GHz
luminosity of the BCG for the B55 and the extended HIFLUGCS samples (Section 7.4).
Red points are $kT>$4 keV clusters. Blue triangles are $kT$=2-4 keV poor
clusters, while green stars are $kT<$2 keV groups. There is a general
trend that strong radio AGN only exist in gas cores with short cooling time.
However, there are three outliers in the B55 sample and seven outliers in the
extended HIFLUGCS sample above $L_{\rm 1.4 GHz}$ of 10$^{24}$ W Hz$^{-1}$ (eight in total, all marked).
All eight BCGs have coronae that are smaller than 10 kpc in radius.
This again shows that BCGs with luminous radio AGN either host a
large cool core or a small corona.
Those in HIFLUGCS sample were considered as noncool core clusters by Mittal et al. (2009).
BCGs of weak cool cores and noncool cores ($t_{\rm cooling, 10 kpc} >$ 2 Gyr)
without a corona only have weak radio AGN. Coma have BCG coronae (Vikhlinin et al. 2001)
and the shown BCG is NGC~4874.
}
\end{figure}

\end{document}